\documentclass[peerreview]{IEEEtran}

\usepackage{cite}

\ifCLASSINFOpdf
\else
\fi
\usepackage[cmex10]{amsmath} 
\usepackage{amsthm,amssymb}

\usepackage{array}

\usepackage{mdwmath}
\usepackage{mdwtab}

\usepackage[tight,footnotesize,it]{subfigure}

\usepackage{xcolor,color,graphicx}
\usepackage[bookmarks,colorlinks,
			pdfauthor={Guido Carlo Ferrante},%
			pdftitle={TranIT}]{hyperref}%
\definecolor{CiteColor}{rgb}{0.99,0.22,.00}
\definecolor{LinkColor}{rgb}{0.22,0.11,.99}
						\hypersetup{colorlinks,%
						citecolor= green!66!black!100,%
						filecolor= blue,%
						linkcolor= blue,%
						urlcolor= blue}

\usepackage{enumerate}
\usepackage{stmaryrd}
\newcommand{\bset}[1]{[{#1}]}
\def\clap#1{\hbox to 0pt{\hss#1\hss}}

\usepackage{etoolbox}

\theoremstyle{remark}
\newtheorem{theorem}{Theorem} 

\newtheorem{corollary}{Corollary}
\newtheorem{lemma}{Lemma}

\newtheorem{definition}{Definition}

\newtheorem{rem}{Remark}

\AtEndEnvironment{definition}{\null\hfill\ensuremath{\blacksquare}}%
\AtEndEnvironment{assumption}{\null\hfill\qedsymbol}%
\AtEndEnvironment{rem}{\null\hfill\qedsymbol}%
\AtEndEnvironment{example}{\null\hfill\qedsymbol}%

\usepackage{braket}
\usepackage{booktabs}
\let\Gamma\varGamma
\let\Delta\varDelta
\let\Theta\varTheta
\let\Lambda\varLambda
\let\Xi\varXi
\let\Pi\varPi
\let\Sigma\varSigma
\let\Upsilon\varUpsilon
\let\Phi\varPhi
\let\Psi\varPsi
\let\Omega\varOmega

\usepackage{caption}
\usepackage{mathrsfs}
\def\indicator{\text{\usefont{U}{bbold}{m}{n}{1}}}
\usepackage{bm} 
\newcommand{\bs}[1]{\bm{#1}}

\let\CMcal\CMcalText
\usepackage[bb=boondox]{mathalfa}
\usepackage{rotating,ragged2e}

\usepackage{enumerate}

\usepackage{multirow}
\newcommand{\numberset}[1]{\mathbb{#1}} 
\newcommand{\C}{\numberset{C}}
\newcommand{\R}{\numberset{R}}

\newcommand{\N}{\CMcal{N}_{\msf{0}}}
\newcommand{\En}{\CMcal{E}}

\newcommand{\Norm}[2]{\ensuremath\text{\usefont{OMS}{cmsy}{m}{n}{N}}(#1,#2)}

\newcommand{\CNorm}[2]{\ensuremath\text{\usefont{OMS}{cmsy}{m}{n}{CN}}(#1,#2)}
\newcommand{\Eb}{{\slim}\CMcal{E}_{b}}
\newcommand{\EbN}{\frac{\CMcal{E}_{b}}{\N}}
\DeclareMathOperator{\Expectation}{\mathbb{E}\!}
\newcommand{\Pois}[1]{\textup{Poisson}(#1)}
\newcommand{\E}[1]{\Expectation\left[\,#1\,\right]}
\newcommand{\Einline}[1]{\Expectation\; [\,#1\,]}
\newcommand{\Var}[1]{\textup{Var}\!\left[\,#1\,\right]}
\newcommand{\Varinline}[1]{\textup{Var}[\,#1\,]}

\usepackage{xargs}
\newcommandx{\yaHelper}[2][1=\empty]{%
\ifthenelse{\equal{#1}{\empty}}%
  { \ensuremath{ \scriptstyle{ #2 } } } 
  { \raisebox{ #1 }[0pt][0pt]{ \ensuremath{ \scriptstyle{ #2 } } } }  
}
\newcommandx{\yrightarrow}[4][1=\empty, 2=\empty, 4=\empty, usedefault=@]{%
  \ifthenelse{\equal{#2}{\empty}}
  { \xrightarrow{ \protect{ \yaHelper[ #4 ]{ #3 } } } } 
  { \xrightarrow[ \protect{ \yaHelper[ #2 ]{ #1 } } ]{ \protect{ \yaHelper[ #4 ]{ #3 } } } } 
}
\def\toprinline{\,{\yrightarrow{\textit{p}}[-0.2ex]}\,}
\def\todistrinline{\,{\yrightarrow{\textit{d}}[-0.3ex]}\,}
\def\toasinline{\,{\yrightarrow{\msf{a.s.}}[-0.3ex]}\,}
\def\toas{\;{\yrightarrow{\; \msf{a.s.}\; }[-0.3ex]}\;}
\def\topr{\;{\yrightarrow{\; \textit{p}\; }[-0.2ex]}\;}

\newcommand{\Probinline}[1]{\mathbb{P}(\,#1\,)}

\def\Jidx{\mathcal{J}}

\newcommand{\intint}[2]{{[}#1{{\slim}:{\slim}}#2{]}}

\DeclareMathOperator{\Ker}{Ker}
\DeclareMathOperator{\Imag}{Im}
\DeclareMathOperator{\rank}{rank}

\DeclareMathOperator{\Tr}{\msf{tr}}

\def\herm{\dag}
\def\t{{\msf{T}\!}}

\DeclareMathOperator{\Log}{log_{10}}

\def\slim{\hspace{0.2ex}}
\def\nslim{\hspace{-0.2ex}}
\def\ie{\textit{i.e.}}
\def\eg{\textit{e.g.} }
\newcommand{\msf}[1]{\mathsf{#1}}
\def\cov{\bs{\Sigma}}

\let\geq\geqslant
\let\leq\leqslant

\makeatletter
\newcommand*{\eqdef}{:=}
\newcommand*{\defeq}{=:}

\newlength\ploth 
\newlength\plotw 
\begin{document}
%
\title{Spectral Efficiency of Random Time-Hopping CDMA}

%
%
%

\author{Guido Carlo Ferrante,~\IEEEmembership{Student Member,~IEEE,}
        and Maria-Gabriella Di Benedetto,~\IEEEmembership{Senior Member,~IEEE}%
\thanks{Manuscript received November 18, 2013; revised November 25, 2014; September 27, 2015. The associate editor coordinating the review was Prof. R. Sundaresan.}
\thanks{G. C. Ferrante was with the Department
of Information Engineering, Electronics and Telecommunications, Sapienza University of Rome, 00184 Rome, Italy, %
and the Department of Telecommunications, CentraleSup\'elec, 91192 Gif-sur-Yvette, France. He is now with Singapore University of Technology and Design, Singapore, and Massachusetts Institute of Technology, MA, within the SUTD-MIT Postdoctoral Programme.}%
\thanks{M.-G. Di Benedetto is with the Department
of Information Engineering, Electronics and Telecommunications, Sapienza University of Rome, 00184 Rome, Italy.}%
} 

%
%

\markboth{Submitted to IEEE Transactions On Information Theory}
{}
%



\maketitle
\begin{abstract}
Traditionally paired with impulsive communications, Time-Hopping CDMA (TH-CDMA) is a multiple access technique that separates users in time by coding their transmissions into pulses occupying a subset of $N_\msf{s}$ chips out of the total $N$ included in a symbol period, in contrast with traditional Direct-Sequence CDMA (DS-CDMA) where $N_\msf{s}=N$. This work analyzes TH-CDMA with random spreading, by determining whether peculiar theoretical limits are identifiable, with both optimal and sub-optimal receiver structures, in particular in the archetypal case of sparse spreading, that is, $N_\msf{s}=1$. Results indicate that TH-CDMA has a fundamentally different behavior than DS-CDMA, where the crucial role played by energy concentration, typical of time-hopping, directly relates with its intrinsic ``uneven'' use of degrees of freedom.

 
\end{abstract}


%

\section{Introduction}

\IEEEPARstart{W}{hile} Direct-Sequence CDMA (DS-CDMA) is widely adopted and thoroughly analyzed in the literature, Time-Hopping CDMA (TH-CDMA) remains a niche subject, often associated with impulsive ultra-wideband communications; as such, it has been poorly investigated in its information-theoretical limits. This paper attempts to fill the gap, by addressing a reference basic case of synchronous, power-controlled systems, with random hopping. 

Time-hopping systems transmit pulses over a subset of chips of cardinality $N_\msf{s}$ out of the $N$ chips composing a symbol period. In contrast to common DS-CDMA, where each chip carries one pulse, and therefore the number of transmitted pulses per symbol is equal to the number of chips, \ie, $N_\msf{s}=N$, time-hopping signals may contain much fewer nonzero chips, in which pulses are effectively used, \ie, $N_\msf{s}\ll N$. Asymptotically, as the number of chips in a symbol period grows, the fraction of filled-in chips in TH vanishes, \ie, $N_\msf{s}/N\to 0$, making TH intrinsically different, the performance of which cannot be derived from that of DS. 
%
TH vs. DS reflect ``sparse'' vs. ``dense'' spreading, where  degrees of freedom, that is, dimensions of the signal space, are ``unevenly'' vs. ``evenly''  used \cite{MedGal:2002, GalMed:1997, TelTse:2000, PorTseNac:2007, BigPro:1998}. In our setting, as further explored in the paper, degrees of freedom coincide with chips; while DS ``evenly'' uses chips, TH adopts the opposite strategy. In this regard, it is evident that DS and TH represent two contrasting approaches, that will be compared, under the assumption of same bandwidth and same per-symbol energy, in terms of spectral efficiency.

\subsection{Related Work}
Although we will show that there exist peculiar theoretical limits for TH-CDMA, their derivation can be carried out within the framework developed by Verd\'u and Shamai \cite{VerSha:1999} and Shamai and Verd\'u \cite{ShaVer:2001}, providing a methodology that is valid for investigating general CDMA with random spreading in the so-called \textit{large-system limit} (LSL), where $K\to\infty$, $N\to\infty$, while $K/N\to\beta$ finite; in particular, \cite{VerSha:1999} provides expressions of spectral efficiency for DS power-controlled systems using optimum as well as linear receivers, while \cite{ShaVer:2001} removes the power-control assumption and introduces fading. Other seminal contributions towards the understanding of random DS-CDMA, although limited to linear receivers, are those of Tse and Hanly \cite{TseHan:1999}, and Tse and Zeitouni \cite{TseZei:2000}. Aside from DS-CDMA, the same framework is aptly used for analyzing other CDMA channels, such as multi-carrier CDMA \cite{TulLiVer:2005}. 

The analysis of optimum decoders relies, in general, on the study of the eigenvalue distribution of random matrices describing random spreading. Consolidated results on the statistical distribution of such eigenvalues of DS matrices \cite{MarPas:1967} form the basis for a tractable analysis of theoretical limits in terms of spectral efficiency. In particular, it is shown in \cite{VerSha:1999} that a fixed loss, that depends upon the load, \ie, the ratio $\beta$ between the number of users $K$ and chips $N$, is incurred with DS vs. orthogonal multiple-access. This loss becomes negligible with optimum decoding when $\beta \gg 1$ while, for $\beta \ll 1$, even a linear receiver such as MMSE is sufficient for achieving this negligible loss; however, this is no longer the case for simpler linear receivers, such as the single-user matched filter (SUMF), that is shown to be limited in spectral efficiency at high SNR. 
As a matter of fact, the above findings on spectral efficiency of DS-CDMA strongly depend on the statistical properties of the eigenvalue distribution, and as such on the cross-correlation properties of the spreading sequences. By changing the spreading strategy from DS to TH, it can be predicted that different theoretical limits will hold, as will be investigated below. In particular, TH matrices, as rigorously defined in this paper, are a special subset of sparse matrices, where the number of nonzero entries is small compared to the total number of elements. 
Previous work on sparse CDMA relies on non-rigorous derivations based on replica methods, 
which are analytical tools borrowed from statistical physics, as pioneered by Tanaka \cite{Tan:2002}, who provides an expression of capacity when inputs are binary. Montanari and Tse \cite{MonTse:2006} propose a rigorous argument for $N_\msf{s}\to\infty$, proving Tanaka's formula, that is valid up to a maximum load, called \textit{spinodal} $(\beta_\mathrm{s}\approx 1.49)$. Above the spinodal load, Tanaka's formula remains unproved. Binary sparse CDMA is also analyzed in terms of detection algorithms, in particular in the so-called belief propagation \cite{MonTse:2006,GuoWan:2008,TanOka:2005}. More recently, capacity bounds for binary sparse CDMA are derived in \cite{Alietal:2009,KorMac:2010}. Still relying on replica methods, \cite{RaySaa:2007} and \cite{YosTan:2006} analyze two different regimes, where $N_\msf{s}$ is either finite or random with fixed mean.


\subsection{Main Contribution and Novelties}
The main contribution of the present work is to provide rigorous information-theoretical limits of time-hopping communications, by inscribing this particular time-domain sparse multiple access scheme into the random matrix framework developed by Verd\'u and Shamai in \cite{VerSha:1999}, for analyzing random spreading. The present analysis allows comparing TH vs. DS with same energy per symbol and same bandwidth constraints, and, therefore, highlights the effect of the energy ``concentration,'' that is typical of TH. A first contribution consists in providing a closed form expression for spectral efficiency of TH with optimum decoding when $N_\msf{s}=1$. A second contribution is to prove that the spectral efficiency formula for a bank of single-user matched filter obtained by Verd\'u and Shamai in \cite{VerSha:1999} for DS systems $(N_\msf{s}=N)$ remains valid if $N_\msf{s}\to\infty$, $N\to\infty$, and $N_\msf{s}/N\to\alpha\in(0,1]$. A third contribution is to provide understanding of when TH performs better than DS.

Based on the above contributions,  
we are able to present a novel interpretation of TH-CDMA against DS-CDMA, that offers a better understanding of the effect of sparsity in time. 

\subsection{Paper Organization}
The paper is organized as follows: in Section~\ref{sec:ref} we describe the model of the synchronous CDMA channel adopted throughout the paper, and particularized to the special case of time-hopping. Section~\ref{sec:synchnonimp} contains the derivation of spectral efficiency of TH-CDMA for different receiver structures, in particular optimum decoding as well as sub-optimal linear receivers, and a comparison with traditional DS-CDMA limits \cite{VerSha:1999}. Conclusions are drawn in Section~\ref{sec:conc}. 

\subsection{Notations}
Boldface letters denotes vectors when lowercase, and matrices when uppercase. The $i$th vector of the standard basis of $\R^n$ is denoted by $\bs{e}_i$. The $j$th element of a vector $\bs{v}$ is denoted by $[\bs{v}]_j$ and the Kronecker delta is denoted by $\delta_{ij}$, hence, for example, $[\bs{e}_i]_j=\delta_{ij}$. The set of integers $\{1,\dotsc,N\}$ is denoted by $\bset{N}$. The cardinality of a set $A$ is denoted by $|A|$. Gaussian distributions with mean $\mu$ and variance $\sigma^2$ are indicated by $\Norm{\mu}{\sigma^2}$ and $\CNorm{\mu}{\sigma^2}$ when referring to real and complex random variables (RVs), respectively; we denote by $\Norm{x;\mu}{\sigma^2}$ and $\CNorm{z;\mu}{\sigma^2}$ their PDFs, by expliciting the argument $x\in\R$ and $z\in\C$, respectively. The notation is straightforwardly generalized to multivariate distributions. Binomial distribution when the number of trials is $n$ and the success probability is $p$ is denoted by $\textup{Binomial}(n,p)$. Poisson distribution with mean $\beta$ is denoted by $\Pois{\beta}$. Convergence in distribution of a sequence of RVs $(X_n)_{n\geq0}$ to $X$ with distribution $P_X$ is denoted $X_n\todistrinline P_X$, while convergence in probability is denoted $X_n\topr X$. The differential entropy of $X$ is denoted either $h(X)$ or $h(P_X)$, where $P_X$ is the distribution of $X$.

\section{Reference Model}\label{sec:ref}


\begin{figure*}[tb]
\centering
\subfigure[DS-CDMA: $N=8$.]{
\includegraphics{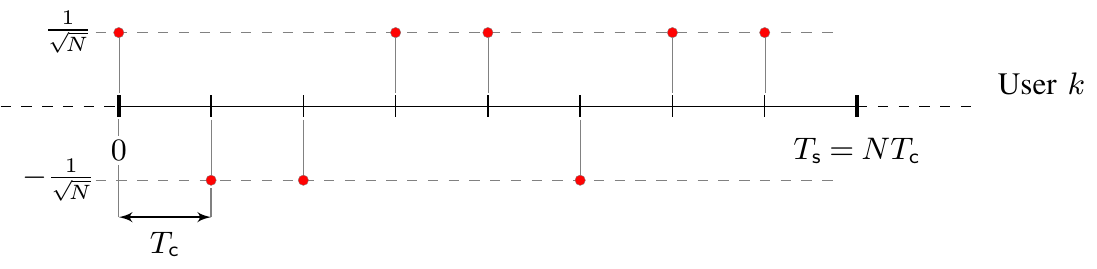}
}
\subfigure[TH-CDMA: $N=8$, $N_\msf{s}=4$, $N_\msf{h}=2$.]{
\includegraphics{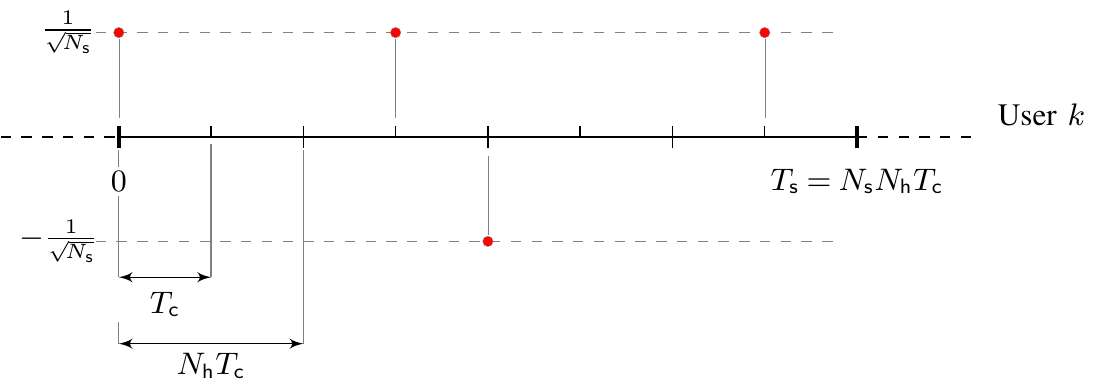}
}
\caption{DS-CDMA vs. TH-CDMA time-axis structure. The symbol period is divided into $N=8$ chips in both figures. In DS-CDMA (Fig.~\ref{fig:refsynchTH-CDMA}\textit{a}), each chip is used for transmitting one pulse, hence eight pulses are transmitted per symbol period. The signature sequence shown on figure is $\bs{s}_k=\frac{1}{\sqrt{8}}[1,-1,-1,1,1,-1,1,1]^\msf{T}$. In TH-CDMA (Fig.~\ref{fig:refsynchTH-CDMA}\textit{b}) the symbol period is divided into $N_\msf{s}=4$ subgroups of $N_\msf{h}=2$ contiguous chips: one pulse only per subgroup is transmitted, that is four pulses in total. The signature sequence shown on figure is $\bs{s}_k=\frac{1}{2}[1,0,0,1,-1,0,0,1]^\msf{T}$. Total energy per symbol is identical in both cases, and equal to one.}
\label{fig:refsynchTH-CDMA}
\end{figure*}

Consider the traditional discrete complex-valued code-division multiple access channel model without fading and with power control \cite{Verdu:1998,ShaVer:2001}: 
\begin{equation}
\label{eq:discrsynch}
\bs{y} = \bs{S}\bs{b}+\bs{n},
\end{equation}
where $\bs{y}\in\C_N$ is the received signal vector with $N$ chips, $\bs{b}{\negthinspace}={\negthinspace}[\msf{b}_1,\dotsc,\msf{b}_K]^\msf{T}\in\C^K$ is the vector of symbols transmitted by the $K$ users, $\bs{S}=[\bs{s}_1,\cdots,\bs{s}_K]\in\R^{N\times K}$ is the spreading matrix, where its $k$th column is the spreading sequence $\bs{s}_k$ of the $k$th user, and $\bs{n}{\in}\C_N$ is a circularly symmetric Gaussian vector with zero mean and covariance $\N\bs{I}$. Here, spreading sequences $\{\bs{s}_k\}_{k=1}^K$ have unit norm, $\|\bs{s}_k\|=1$, and users are subject to a common power constraint, %
$\E{ |\msf{b}_k|^2 } \leq \CMcal{E}$, $1 \leq k \leq K$. It is assumed, as is common (\eg \cite{Ver:2002}), that symbols of different users are independent; this, with the power constraint, leads to:
\begin{equation}
\label{eq:pcon} \Expectation\;[{\slim}\bs{b}\bs{b}^\herm{\slim}] = \CMcal{E} \bs{I}.
\end{equation}
Different CDMA systems can be studied by specifying the spreading matrix: in particular, random CDMA systems are described by random spreading matrices. For example, in random DS-CDMA, spreading sequences may be binary sequences, with elements typically modeled as Bernoulli random variables, $[\bs{s}_k]_i\in\{ -1/\sqrt{N}, +1/\sqrt{N} \}$, $i=1,\dotsc,N$, drawn with equal probability, or spherical sequences, with $\bs{s}_k$ a unitarily invariant unit-norm vector \cite{VerSha:1999}. 

In order to cast TH-CDMA in the model described by eq. \eqref{eq:discrsynch}, let $N=N_\msf{s}\cdot N_\msf{h}$, \ie, the $N$ chips are divided into $N_\msf{s}$ subgroups, and each of these $N_\msf{s}$ subgroups is made of $N_\msf{h}$ contiguous chips. The generic element of a signature sequence can take values in $\{\-1/\sqrt{N_\msf{s}}, 0, +1/\sqrt{N_\msf{s}}\}$, %
and the structure of each sequence $\bs{s}_k$ is such that there is one and only one nonzero element within each of the $N_{\msf{s}}$ subgroups. Therefore, the number of nonzero elements of each signature sequence is fixed to $N_\msf{s}$. We formally introduce the new structure of spreading sequences by the two following definitions.
\medskip
\begin{definition}[\textit{Sparse vector}] A vector $\bs{s}=[s_1,\dotsc,s_{N}]^\t\in\C_N$ is $S$-sparse if the subset of its nonzero elements has cardinality $S$, \ie, $|\{ s_i\neq 0\colon i=1,\dotsc,N \}|=S$. \end{definition}\medskip
\begin{definition}[\textit{$(N_\msf{s},N_\msf{h})$-sequence, TH and DS sequences and matrices}] \label{def:THseq} A vector $\bs{s}=[s_1,\dotsc,s_{N}]^\t\in\C^{N\times 1}$ is a $(N_\msf{s},N_\msf{h})$-sequence when: 
\begin{enumerate}
\item $N=N_\msf{s}\cdot N_\msf{h}$, with $N_\msf{s}\in\mathbb{N}$ and $N_\msf{h}\in\mathbb{N}$; 
\item for all $1 \leq m \leq N_\msf{s}$, the vector $[s_{1+(m-1)N_\msf{h}},\dotsc,s_{m N_\msf{h}}]^\t$ is $1$-sparse, where the nonzero element is either $-1/\sqrt{N_\msf{s}}$ or $1/\sqrt{N_\msf{s}}$ with equal probability, and is drawn uniformly at random.
\end{enumerate}
A $(N_\msf{s},N_\msf{h})$-sequence with $N_\msf{s}<N$ is a Time-Hopping (TH) sequence; the special case $N_\msf{s}=N$, \ie, $(N,1)$-sequences corresponds to binary DS sequences, that will be referred to below simply as DS sequences. A matrix $\bs{S}$ is called TH vs. DS matrix when its columns correspond to TH vs. DS sequences. The set of all possible TH vs. DS matrices is indicated as TH vs. DS ensemble.
\end{definition}\medskip

Figure~\ref{fig:refsynchTH-CDMA} shows the organization of the time axis for DS-CDMA (Fig.~\ref{fig:refsynchTH-CDMA}\textit{a}) and compares this time pattern against TH-CDMA (Fig.~\ref{fig:refsynchTH-CDMA}\textit{b}). Note that, for $N_\msf{s}{\negthinspace}={\negthinspace}N$, TH-CDMA reduces to DS-CDMA.  The unit-norm assumption on spreading sequences implies that the ensuing comparison of TH-CDMA vs. DS-CDMA is drawn under the constraint of same energy per sequence. Note that the $N_\msf{s}=1$ case models a strategy of maximum energy concentration in time, while maximum energy spreading in time corresponds to making $N_\msf{s}=N$, as in DS. Also note that the two systems operate under same bandwidth constraint.

\section{Spectral Efficiency of TH-CDMA}\label{sec:synchnonimp}
In this section, spectral efficiency of TH-CDMA is derived for different receiver structures, and compared against consolidated results for DS-CDMA \cite{VerSha:1999}.
 
The section is organized as follows: we first analyze the case of optimum decoding (Section~\ref{sub:syn}), then proceed to linear receivers in sections \ref{sub:sumf} and \ref{sub:deco} for single-user matched filters (SUMF), and decorrelator/MMSE receivers, respectively. Finally, Section~\ref{sub:synopsis} contains a synposis.

\subsection{Optimum decoding} 
\label{sub:syn}

\subsubsection{Theoretical framework} 
In general terms, a key performance measure in the coded regime is spectral efficiency $C^\msf{opt}$ (b/s/Hz) as a function of either signal-to-noise ratio $\gamma$ or energy per bit $\Eb$-to-noise-$\N$, $\Eb/\N$. 

Referring to model of eq.~\eqref{eq:discrsynch}, where the dimension of the observed process is $N$, spectral efficiency is indicated as $C^\msf{opt}_N(\gamma)$ and is the maximum mutual information between $\bs{b}$ and $\bs{y}$ knowing $\bs{S}$ over distributions of $\bs{b}$, normalized to $N$. Under constraint \eqref{eq:pcon}, %
$C^\msf{opt}_N(\gamma)$ (b/s/Hz) is achieved with Gaussian distributed $\bs{b}$, and it is expressed by \cite{Tel:1999,VerSha:1999,TulVer:2004, Tul:2009, CouDeb:2011}: 
\begin{equation}
\label{eq:sesynch}
C^\msf{opt}_N(\gamma) = \frac{1}{N} \log_{2}\det(\bs{I}+\gamma{\slim} \bs{S}\bs{S}^\t),
\end{equation}
where noise has covariance $\cov_{\bs{n}}=\N \bs{I}$ and $\gamma$ is given by \cite{Ver:2002}:
\begin{equation}
\label{eq:snrsynch}
\gamma \eqdef \frac{\frac{1}{K}\E{\|\bs{b}\|^2}}{\frac{1}{N}\E{\|\bs{n}\|^2}} = \frac{\frac{1}{K}{\negthinspace}\cdot b\Eb}{\frac{1}{N}{\negthinspace} \cdot N \N} = \frac{1}{\beta}\cdot \frac{b}{N}\cdot \EbN = \frac{1}{\beta}\cdot C^\msf{opt}_N\cdot \eta,
\end{equation}
where $\beta\eqdef K/N$ is the \textit{load}, $\eta\eqdef \Eb/\N$, $b$ is the number of bits encoded in $\bs{b}$ for a capacity-achieving system, and therefore $b/N$ coincides with spectral efficiency $C^\msf{opt}_N$ of eq.~\eqref{eq:sesynch}. Since $N$ is equal to the number of possible complex dimensions, spectral efficiency can, therefore, be interpreted as the maximum number of bits per each complex dimension. Note that the number of complex dimensions coincides in our setting with the degrees of freedom of the system, that is, with the dimension of the observed signal space.

\smallskip

Eq.~\eqref{eq:sesynch} can be equivalently rewritten in terms of the set of eigenvalues $\{ \lambda_n(\bs{S}\bs{S}^{\t})\colon n=1,\dotsc,N \}$ of the Gram matrix $\bs{S}\bs{S}^{\t}$ as follows:
\begin{equation}
\label{eq:seintegrala}
C^\msf{opt}_{N}(\gamma) = \frac{1}{N} {\negthinspace}\sum_{n=1}^N \log_{2}(1+\lambda_n \gamma)=\int_0^\infty \log_{2}(1+\lambda\gamma)\, d{\msf{F}^{\bs{S}\bs{S}^\t}_{\!N}\!}(\lambda), 
\end{equation}
where ${\msf{F}^{\bs{S}\bs{S}^\t}_{\!N}\!}(x)$ is the so called \textit{empirical spectral distribution} (ESD) defined as \cite{TulVer:2004}:
\begin{equation}\label{eq:esddef} 
{\msf{F}^{\bs{S}\bs{S}^\t}_{\!N}\!}(x) \eqdef \frac{1}{N} \sum_{n=1}^N \indicator\{\lambda_n(\bs{S}\bs{S}^\t)\leq x\}, 
\end{equation}
that counts the fraction of eigenvalues of $\bs{S}\bs{S}^\t$ not larger than $x$. Being $\bs{S}$ random, so is the function ${\msf{F}^{\bs{S}\bs{S}^\t}_{\!N}\!}$. The limit distribution of the sequence $\{{\msf{F}^{\bs{S}\bs{S}^\t}_{\!N}\!}\colon N\geq 1\}$, when it exists, is called \textit{limiting spectral distribution} (LSD) and denoted $\msf{F}$; it turns out that $\msf{F}$, differently from ${\msf{F}^{\bs{S}\bs{S}^\t}_{\!N}\!}$, is usually nonrandom \cite{BaiSil:2009}. In particular, the regime of interest, referred to as \textit{large-system limit} (LSL), is that of both $N\to\infty$ and $K\to\infty$ while keeping $K/N\to\beta$ finite. In the LSL, spectral efficiency $C^\msf{opt}_{N}(\gamma)$, that is a random variable, may converge to
\begin{align}
\label{eq:cinfsynch}
C^\msf{opt}(\gamma) \eqdef \int_0^\infty \log_{2}(1+\lambda\gamma)\, d\msf{F}(\lambda),
\end{align}
where the convergence mode has to be specified. In general, convergence of ${\msf{F}^{\bs{S}\bs{S}^\t}_{\!N}\!}$ to $\msf{F}$ does not imply convergence of $C^\msf{opt}_{N}$ to $C^\msf{opt}$, that must be proved.

Therefore, finding the spectral efficiency of CDMA systems with random spreading in the LSL regime reduces to finding the LSD $\msf{F}(\lambda)$, that depends on the spreading sequence family only; hence, in the rest of this section, we find the LSD of TH-CDMA with $N_{\msf{s}}=1$, which corresponds to a maximum energy concentration in time, as well as asymptotic behaviors of TH-CDMA systems with generic $N_{\msf{s}}$.

\medskip

\subsubsection{LSD and spectral efficiency of TH-CDMA systems with $N_\msf{s}=1$} 

While for DS-CDMA, spectral efficiency can be computed directly from Mar\u{c}enko and Pastur result on the ESD of matrices with i.i.d. elements \cite{MarPas:1967}, it appears that no analog result is available for neither TH-CDMA matrices nor dual matrices describing frequency-hopping.

We hereby derive the LSD and properties of the ESD of synchronous TH-CDMA when $N_\msf{s}=1$ by means of the method of moments. 
In Theorem~\ref{theo:Nsonemomentsna} we derive properties of the $L$th moment of the ESD ${\msf{F}^{\bs{S}\bs{S}^\t}_{\!N}\!}$ with $N_\msf{s}=1$, denoted by:
\[ m_L\eqdef \frac{1}{N}\Tr(\bs{S}\bs{S}^\t)^L = \int_0^\infty \lambda^L\,d {\msf{F}^{\bs{S}\bs{S}^\t}_{\!N}\!}(\lambda), \]
in particular a closed form expression of $\E{m_L}$ for TH matrices with $N_\msf{s}=1$ for finite $K$ and $N$, and we prove convergence in probability to moments of a Poisson distribution with mean $\beta$ in the LSL.

\begin{theorem}
\label{theo:Nsonemomentsna}
Suppose that $\bs{S}\in\R^{N\times \beta N}$ is a time-hopping matrix with $N_\msf{s}=1$. Then, 
in the LSL, $m_L$ converges in probability to the $L$th moment of a Poisson distribution with mean $\beta$, \ie:
\[ m_L {\topr} \bar{m}_L\eqdef \sum_{\ell=1}^L \left\{ \nslim{L \atop \ell} \right\} \beta^\ell, \]
where $\Big\{ \nslim{L \atop \ell} \Big\}$ denotes a Stirling number of the second kind.
\end{theorem}
\IEEEproof{See Appendix~\ref{app:dimfinite}.\hfill$\square$\medskip}

In general, the Carleman condition guarantees that the set of moments uniquely defines the probability distribution. For the sake of completeness, it is verified in Appendix~\ref{app:carleman}. Hence, the set of moments $\{\bar{m}_L\}_{L\geq1}$ uniquely defines the Poisson distribution, and Theorem~\ref{theo:Nsonemomentsna} implies that the LSD is a Poisson law with mean $\beta$:
\smallskip

\begin{corollary} 
\label{theo:Nsone}
Suppose that $\bs{S}\in\R^{N\times \beta N}$ is a time-hopping matrix, as specified in Definition~\ref{def:THseq}, with $N_\msf{s}=1$. Then, the ESD of $\{\bs{S}\bs{S}^\t\colon N\geq 1\}$ converges in probability to the distribution function $\msf{F}$ of a Poisson law with mean $\beta$:
\begin{equation}\label{eq:Nsone}
{\msf{F}^{\bs{S}\bs{S}^\t}_{\!N}\!}(x) {\topr} \msf{F}(x)=\sum_{k\geq 0} \frac{\beta^k e^{-\beta}}{k!} \indicator\{k\leq x\}.
\end{equation}
\end{corollary}



\medskip
%
%
In terms of measures, TH-CDMA is thus characterized by the purely atomic measure given by:
\begin{equation}\label{eq:measth} \mu^\msf{TH} \eqdef \sum_{k\geq 0} \frac{\beta^k e^{-\beta}}{k!}\, \delta_k =\sum_{k\geq 0} \msf{f}_k(\beta) \delta_k,\end{equation}
being $\msf{f}_k(\beta)\eqdef {\beta^k e^{-\beta}}/{k!}$, and $\delta_k$ the point mass distribution, \ie, $\delta_k(A)=1$ if $k\in A$, and $\delta_k(A)=0$ otherwise. Whence, $\msf{F}(x)=\mu^\msf{TH}((-\infty,x])=\mu^\msf{TH}([0,x])$. The above implies peculiar properties of TH-CDMA when compared against DS-CDMA. For convenience, we report here the Mar\u{c}enko-Pastur law, that is the LSD of eigenvalues of DS-CDMA matrices (see Definition~\ref{def:THseq}), which has measure:
\begin{equation}
\label{eq:mplaw}
\mu^{\msf{DS}} = (1-\beta)^{+} \, \delta_0 + \mu^{\msf{DS}}_{\textup{ac}}, 
\end{equation}
where $(x)^{+}\eqdef\max{\negthinspace}\big\{0,x\big\}$, and $\mu^{\msf{DS}}_{\textup{ac}}$ is the absolutely continuous part of $\mu^{\msf{DS}}$ with density (Radon-Nikodym derivative with respect to the Lebesgue measure $m$):
\begin{equation}
\label{eq:mpdensity}
\frac{d\mu^{\msf{DS}}_{\textup{ac}}}{dm}(x) = \frac{1}{2\pi x}\sqrt{-(x-\ell^+)(x-\ell^{-})}\; \indicator\{x\in [\ell^-,\ell^+]\} \defeq \msf{f}^{\msf{MP}}(x),
\end{equation}
where $\ell^{\pm}=(1\pm\sqrt{\beta})^2$. 

\setlength\ploth{0.375\columnwidth} \setlength\plotw{0.45\columnwidth} 
\begin{figure}[t]
\centering
\includegraphics{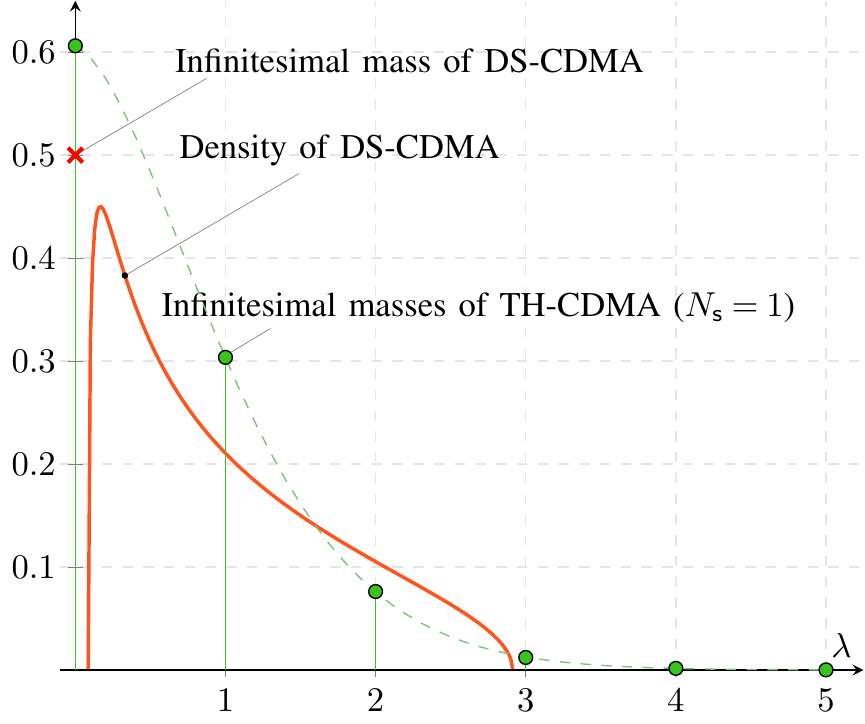}
\caption{Density function of the LSD for DS-CDMA in solide line, and infinitesimal masses of atomic measures of DS-CDMA and TH-CDMA (Poisson law) for $\beta=1/2$ in cross and dots, respectively. DS-CDMA and TH-CDMA with $N_\msf{s}=1$ are governed by Mar\u{c}enko-Pastur and Poisson laws, respectively. }
\label{fig:FIG_MP}
\end{figure}

Fig.~\ref{fig:FIG_MP} shows Mar\u{c}enko-Pastur and Poisson laws for $\beta=1/2$. The Mar\u{c}enko-Pastur law has, in general, an absolute continuous part with probability density function showed in solid line and an atomic part formed by a point mass at the origin showed with a cross at height $1/2$. The Poisson law has a purely atomic (also known as discrete, or counting) measure with point masses at nonnegative integers showed by dots with heights given by $\msf{f}_k(\beta)$ (envelope showed in dashed line). 

\medskip

We use the Poisson LSD to find the spectral efficiency of TH-CDMA with $N_\msf{s}=1$ in the LSL, \ie\  (see eq.~\eqref{eq:seintegrala}):
\begin{equation}\label{eq:capcon}
\int_0^\infty \log_{2}(1+\lambda \gamma) \,d{\msf{F}^{\bs{S}\bs{S}^\t}_{\!N}\!}(\lambda) \;{\topr} \int_0^\infty \log_{2}(1+\lambda \gamma) \,d\msf{F}(\lambda).
\end{equation}
It is important to remark that the above convergence in probability does not follow immediately; in fact, convergence in law does only imply convergence of bounded functionals, but $\log_{2}(1+\lambda \gamma)$ is not bounded on the support of $\msf{F}(\lambda)$. We prove eq.~\eqref{eq:capcon} in Appendix~\ref{app:CntoC}, and thus:
\begin{equation}
\label{eq:sens1}
C^\msf{opt}_{N}(\gamma){\topr}C^\msf{opt}(\gamma) = \sum_{k\geq 0} \frac{\beta^k e^{-\beta}}{k!} \log_{2}(1+k\gamma).
\end{equation}


%
%
%
%
%
%
%
The capacity of a TH-CDMA system with $N_\msf{s}=1$ can be interpreted as follows. 
%
Rewrite eq. \eqref{eq:sens1} as follows:
\begin{align}\label{eq:sens1bis}
C^\msf{opt}(\gamma) = \sum_{k\geq 0} \msf{f}_k(\beta) \, C_k(\gamma),
\end{align}
where $C_k(\gamma)\eqdef \log_{2}(1+k\gamma)$. %
Hence, $C^\msf{opt}(\gamma)$ is a sum of channel capacities $C_k(\gamma)$, $k\in\numberset{N}$, weighted by probabilities $\msf{f}_k(\beta)$. Since $C_k(\gamma)$ is the capacity of a complex AWGN channel with signal-to-noise ratio $k\gamma$, $k\in\numberset{N}$, $C^\msf{opt}(\gamma)$ is equal to the capacity of an infinite set of complex AWGN channels with increasing signal-to-noise ratio $k\gamma$ paired with decreasing probability of being used $\msf{f}_k(\beta)$. Therefore, TH-CDMA has the same behavior of an access scheme that splits the multiaccess channel into independent channels, each corrupted by noise only, with power gain equal to $k$, and excited with probability $\msf{f}_k(\beta)$. Since $\msf{f}_{k}(\beta)$ is also the probability that $k$ signatures have their nonzero element in the same dimension, that is for TH-CDMA associated with the event of waveforms having their pulse over the same chip, for small $\beta$, that is, $\beta\leq 1$, channels with high capacity (for a fixed $\gamma$), that is, with $k\gg 1$, are less frequently used than channels with low capacity; in general, channels with $k$ in a neighborhood of $\beta$ are used most frequently.





One noticeable difference between DS and TH matrices is that in the former the maximum eigenvalue $\lambda_{\textup{max}} \toasinline (1+\sqrt{\beta})^2$ \cite{BaiSil:1998}, and thus also $\lambda_{\textup{max}}{\toprinline}(1+\sqrt{\beta})^2$, while in the latter $\lambda_{\textup{max}}{\toprinline}\infty$. 



Moreover, there exists a nonzero probability $\msf{f}_{0}(\beta)=e^{-\beta}$ such that, also for $\beta>1$, the zero-capacity channel ($k=0$) is excited. This probability, that is the amplitude of the Dirac mass at $\lambda=0$, is equal to $\msf{F}(0)$; it equals the probability that a chip is not chosen by any user or, equivalently, the average fraction of unused chips; and, finally, it equals the high-SNR slope \textit{penalty}, as we will detail below.

\begin{figure}[t]
\centering
\includegraphics{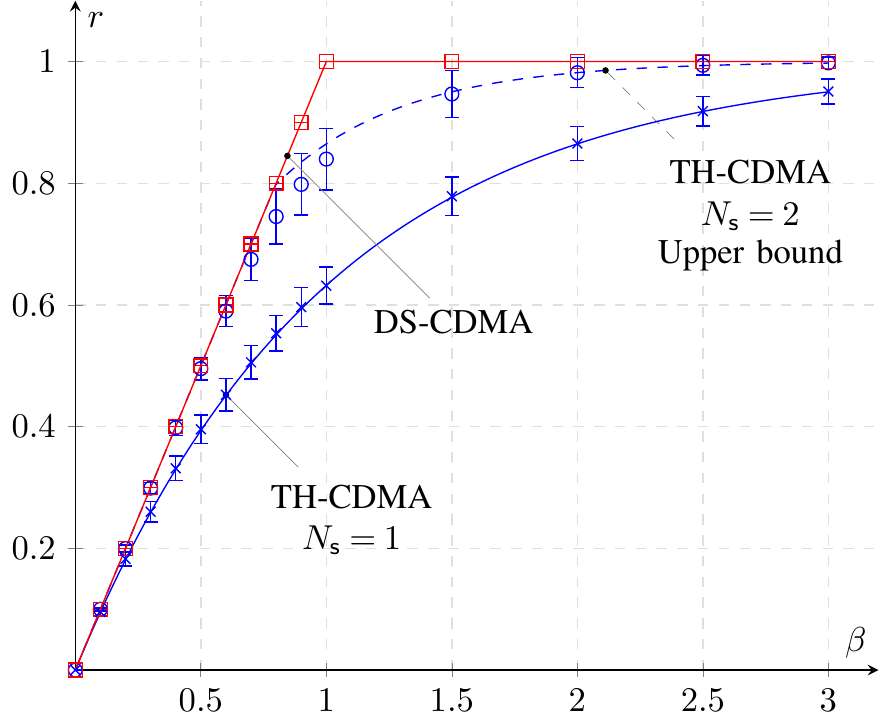}
\caption{Normalized rank ${r}$ (solid lines) vs. load $\beta$. The dashed line represents an upper bound of ${r}$ for TH-CDMA with $N_{\msf{s}}=2$. Crosses, circles, and squares (generally referred to as marks), are obtained by evaluating $\E{ \rank \bs{S}/N }$ by Monte-Carlo simulations of a finite-dimensional system with $N=50$, for TH-CDMA with $N_\msf{s}=1$, TH-CDMA with $N_\msf{s}=2$, and DS-CDMA, respectively. Error bars represent one standard deviation of $\rank \bs{S}/N$.}
\label{fig:figrank}
\end{figure}

It is interesting to analyze the behavior of $r_N\eqdef\rank\bs{S}/N$, that is a random variable for finite $N$.  Figure~\ref{fig:figrank} shows with marks $\bar{r}_N\eqdef \E{r_N}$ for TH-CDMA with $N_\msf{s}=1$ and $N_\msf{s}=2$, and for DS-CDMA, when $N=50$: Monte-Carlo simulations provide point data, represented by marks, with error bars showing one standard deviation of $r_N$. Solid lines represent the limiting value $r$ of $r_N$ as $N\to\infty$. We will show in the below Theorem~\ref{thm:rank1} that, for $N_\msf{s}=1$, $r_N{\toprinline}1-e^{-\beta}$. Almost sure convergence does hold for the Mar\u{c}enko-Pastur law, hence for DS-CDMA one has $r_N \toasinline \min\{1,\beta\}$. For TH-CDMA with increasing $N_{\msf{s}}$, one might expect ${r}$ of TH-CDMA to tend to that of DS-CDMA, also suggested by the behavior of the $N_\msf{s}=2$ case shown on figure. In the general $N_\msf{s}>1$ case, we are able to find the upper bound $r\leq 1-e^{-N_\msf{s}\beta}$ only, holding in probability, that is derived in the below Corollary~\ref{thm:upperbound}, and shown with the dashed line on Fig.~\ref{fig:figrank}. 


\begin{corollary} \label{thm:rank1}
Under the same assumptions of Theorem~\ref{theo:Nsonemomentsna}, it results $r_N\eqdef \frac{1}{N} \rank\bs{S}{\toprinline} 1-e^{-\beta}$.
\end{corollary}
\IEEEproof{When $N_\msf{s}=1$, 
$\rank\bs{S}$ is equal to the number of nonempty rows of $\bs{S}$. Therefore, $r_N{\toprinline}1-e^{-\beta}$. 
%
%
}
\rem{From the definition of ESD, $N{\msf{F}^{\bs{S}\bs{S}^\t}_{\!N}\!}(0)$ is equal to the number of zero eigenvalues, that also provides the dimension of the nullity subspace of $\bs{S}\bs{S}^\t$. Since, from the Rank-Nullity Theorem, $\dim\Ker \bs{S}\bs{S}^\t = N-\dim\Imag \bs{S}\bs{S}^\t = N-\rank \bs{S}\bs{S}^\t = N - \rank\bs{S}$, it results ${\msf{F}^{\bs{S}\bs{S}^\t}_{\!N}\!}(0)=\frac{1}{N} \rank\bs{S}\toprinline 1-e^{-\beta}$.}

\begin{theorem}\label{thm:upperbound} Let $N_\msf{s}\geq 1$. An upper bound to $r$ is given by %
${r} \leq \min\{\beta,1-e^{-N_\msf{s}\beta}\}$, %
which holds in probability.
\end{theorem}
\IEEEproof{Rewrite $\bs{S}$ as follows: $\bs{S}=[\bs{S}_1^\t, \cdots, \bs{S}_{N_\msf{s}}^\t]^\t$, 
where $\{\bs{S}_i\}_{i=1}^{N_\msf{s}}$ are $N_\msf{h}\times K$ matrices, $N_\msf{h}=N/N_\msf{s}$. Using the inequality $\rank(\bs{A}+\bs{B}) \leq \rank\bs{A}+\rank\bs{B}$, %
we can upper bound $\rank\bs{S}$ as follows: $\rank\bs{S}	\leq \; \sum_{i=1}^{N_\msf{s}} \rank\bs{S}_i$. 
Since $\{\bs{S}_i\}_{i=1}^{N_\msf{s}}$ are independent, by Theorem~\ref{thm:rank1} one has ${r}\leq 1-e^{-N_\msf{s}\beta}$ in probability. Moreover, since $\rank\bs{S}/N\leq \min\{1,\beta\}$ surely, we also have $\bar{r}\leq \beta$.
\hfill$\square$}

\setlength\ploth{0.375\columnwidth} \setlength\plotw{0.45\columnwidth} 



\medskip
\subsubsection{Asymptotics}

In the following, spectral efficiency, when expressed as a function of $\eta\eqdef \Eb/\N$, will be indicated by\footnote{In this subsection, we drop the superscript ``$\msf{opt}$'' for ease of notation.} $\msf{C}$ (b/s/Hz), as suggested in \cite{Ver:2002}, rather than $C$ (b/s/Hz), that denotes spectral efficiency as a function of $\gamma$. While an expression of $C$ can be found in terms of the LSD, the same is more difficult for $\msf{C}$, given the nonlinear relation between $C$ and $\msf{C}$: $\msf{C}=C(\eta\msf{C}/\beta)$ (c.f. eq.~\eqref{eq:snrsynch}). 

\setlength\ploth{0.65\columnwidth} \setlength\plotw{0.95\columnwidth} 
\begin{figure*}[t]
\centering
\includegraphics{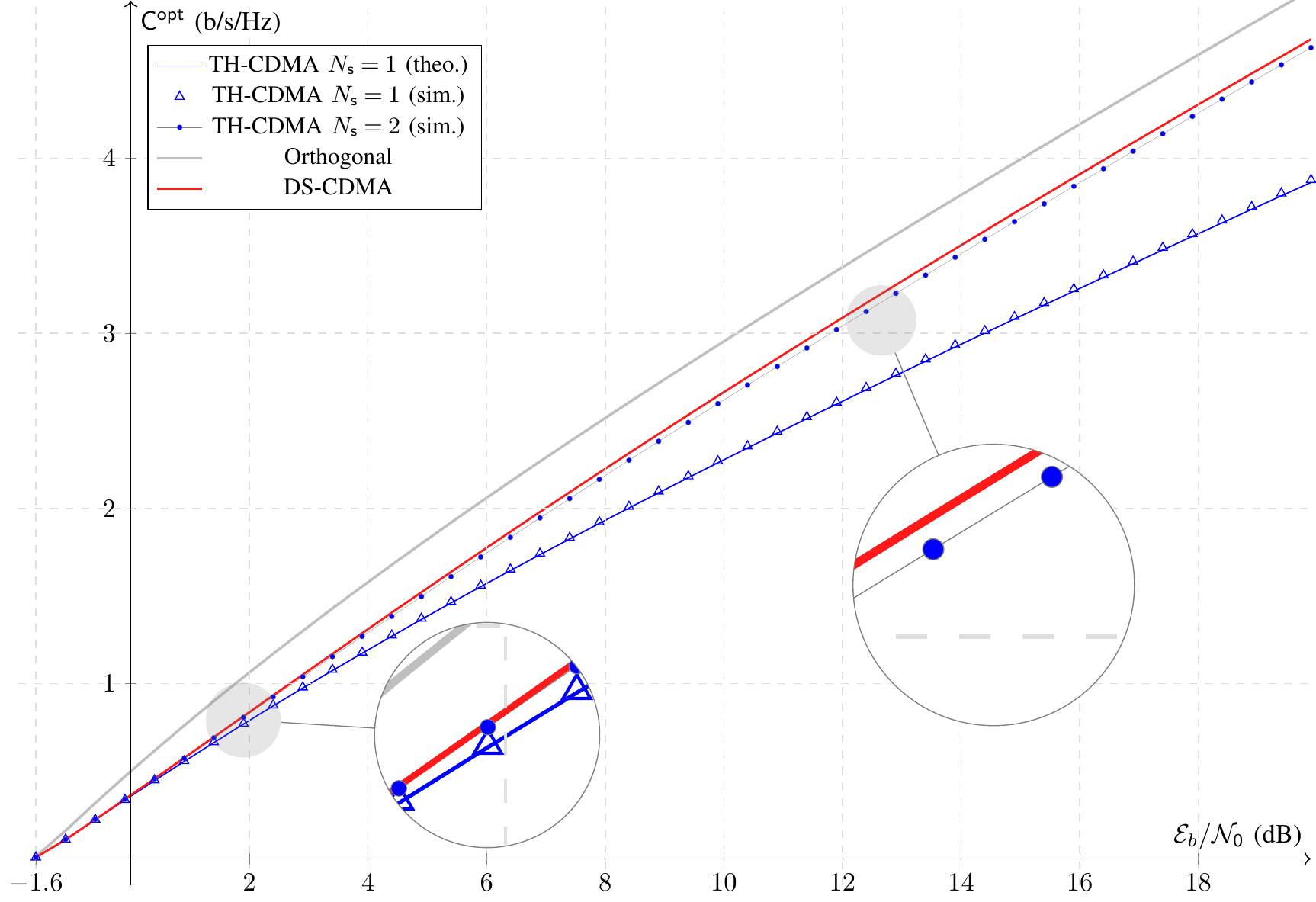}
\caption{Spectral efficiency $\msf{C}^{\msf{opt}}$ (b/s/Hz) of TH-CDMA vs. DS-CDMA with optimum decoding as a function of $\Eb/\N$ (dB) with load $\beta=1/2$. Orthogonal multiple access is reported for comparison (gray solid line). The $N_\msf{s}=1$ TH-CDMA case is plotted for theoretical-values (blue solid line) vs. simulated data (blue triangles). The $N_\msf{s}=2$ TH-CDMA case reports only simulated data (blue dots). DS-CDMA is shown with red solid line. 
Note on figure that TH-CDMA with $N_{\msf{s}}=1$ and $N_{\msf{s}}=2$ have both similar performance as DS in the wideband regime ($\Eb/\N\to\ln2$), while departing from it for high SNR when $N_\msf{s}=1$. Note on figure that the loss incurred with TH drops to a very small value with as early as $N_\msf{s}=2$.
}
\label{fig:fig1}
\end{figure*}

In order to understand the asymptotic behavior of $\msf{C}$ in the low-SNR and high-SNR regimes, \ie, as $\eta\to\eta_{\msf{min}}\eqdef \inf_{\msf{C}>0}\eta(\msf{C})$ and $\eta\to\infty$, respectively,  Shamai and Verd\'u \cite{ShaVer:2001} and Verd\'u \cite{Ver:2002} introduced the following four relevant parameters:
\begin{description}
\item[ $\eta_{\msf{min}}$: ] the \textit{minimum energy per bit over noise level} required for reliable communication; 
\item[ $\CMcal{S}_0$: ] the \textit{wideband slope} (b/s/Hz/(3 dB)); 
\item[ $\CMcal{S}_\infty$: ] the \textit{high-SNR slope} (b/s/Hz/(3 dB)); 
\item[ $\CMcal{L}_\infty$: ] the \textit{high-SNR decibel offset}.
\end{description}

\begin{figure*}[t]
\centering
\setlength\ploth{0.375\columnwidth} \setlength\plotw{0.45\columnwidth}
\includegraphics{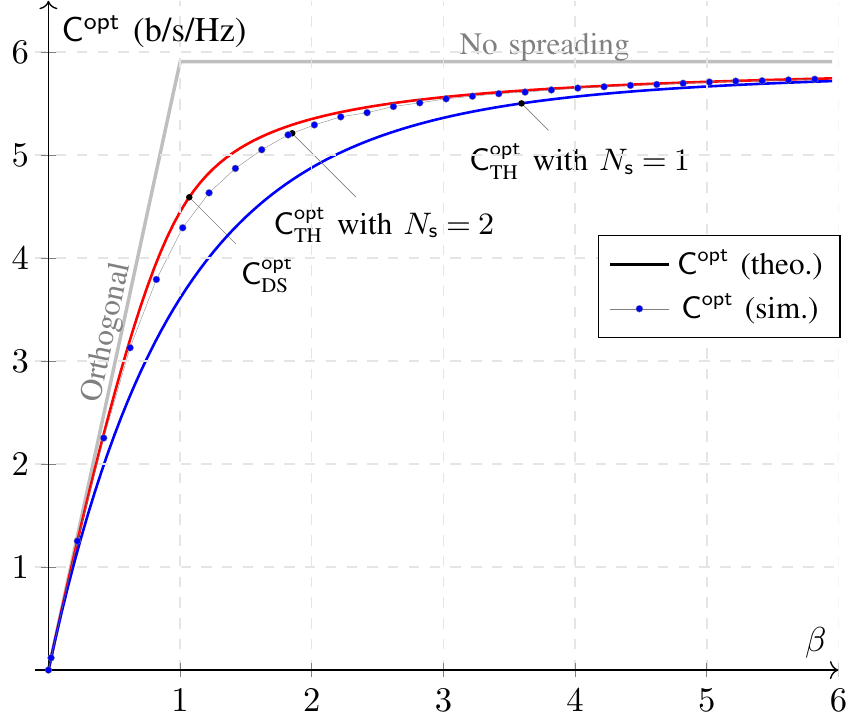}
\caption{Spectral efficiency $\msf{C}^\msf{opt}$ (b/s/Hz) as a function of $\beta$, for $\Eb/\N=10$ dB. DS and TH with $N_\msf{s}=1$ are shown in solid lines, indicating that curves derive from closed form expressions, while values for TH with $N_\msf{s}=2$ are shown in dots, indicating that they derive from simulations. Orthogonal access is also reported for reference (gray solid line).}
\label{fig:Cbeta_0}
\end{figure*}

In our setting, the low-SNR and high-SNR regimes also correspond to $\msf{C}\to0$ (so called wideband regime \cite{Ver:2002}) and $\msf{C}\to\infty$.

The minimum energy-per-bit $\eta_\msf{min}$ and the wideband slope $\CMcal{S}_{0}$ (b/s/Hz/(3 dB)) characterize the affine approximation of $\msf{C}$ vs. $\eta^{\msf{dB}}\eqdef 10\log_{10}\eta$ as $\msf{C}\to0$:
\begin{equation}\label{eq:lowaffine} \eta^\msf{dB} = \eta^\msf{dB}_\msf{min} + \frac{10\Log2}{\CMcal{S}_0}\msf{C}+o(\msf{C}),\quad \msf{C}\to0. \end{equation}
From eq.s~\eqref{eq:lowaffine} and \eqref{eq:snrsynch}, one can find $\eta_\msf{min}$ and $\CMcal{S}_{0}$ as follows:
\begin{align}
\label{eq:lowhowto1}
\eta_\msf{min} & = \lim_{\gamma\downarrow0} \frac{\beta\gamma}{{C}(\gamma)} = \frac{\beta}{C'(0)}=\frac{\beta}{\E{\lambda}}\ln 2,\\
\label{eq:lowhowto2}
\CMcal{S}_0 & = -2\ln2\frac{(C'(0))^2}{C''(0)} = 2\frac{(\E{\lambda})^2}{\E{\lambda^2}},
\end{align}
where the expression in the last equality of both eqs.~\eqref{eq:lowhowto1} and \eqref{eq:lowhowto2} is obtained by differentiating $\log_{2}(1+\lambda\gamma)$ with respect to $\lambda$ under the integral sign.  

The high-SNR slope $\CMcal{S}_{\infty}$ (b/s/Hz/(3 dB)) and high-SNR decibel offset $\CMcal{L}_{\infty}$ characterize the affine approximation of $\msf{C}$ vs. $\eta$ as $\msf{C}\to\infty$:
\begin{equation*}
\eta^{\msf{dB}} = \frac{10\Log 2}{\CMcal{S}_\infty}\,\msf{C}+\CMcal{L}_\infty\,10\Log 2+O(\Log\msf{C}), \quad \msf{C}\to\infty.  
\end{equation*}
Equivalently, the following relation holds in terms of $C$ vs. $\gamma$:
\[ C(\gamma)=\CMcal{S}_\infty \Big[\, \log_{2} (\beta\gamma)-\CMcal{L}_\infty \,\Big]+o(1),\quad \gamma\to\infty, \]
from which $\CMcal{S}_\infty$ and $\CMcal{L}_\infty$ are derived as:
\begin{align}
\label{eq:highhowto1}
\CMcal{S}_\infty&=\lim_{\eta\uparrow\infty} \frac{\msf{C}(\eta)}{\ln \eta}\ln 2=\lim_{\eta\uparrow\infty} \eta\,\msf{C}'(\eta)\ln 2=\lim_{\gamma\uparrow\infty} \gamma C'(\gamma)\ln 2 =1-\msf{F}(0) \\
\label{eq:highhowto2}
\CMcal{L}_\infty & = \log_{2}\beta + \lim_{\gamma\uparrow\infty} \left[ \log_{2}\gamma - \frac{C(\gamma)}{\CMcal{S}_\infty}  \right],
\end{align}
where the last equality in eq.~\eqref{eq:highhowto1} is obtained by differentiating $\log_{2}(1+\lambda\gamma)$ with respect to $\gamma$ and applying the dominated convergence theorem to pass the limit under the integral sign. 

For TH-CDMA with $N_{\msf{s}}=1$, it can be shown by direct computations that the four above parameters are given by:
\begin{align}
\label{eq:Ns1low1}
\eta_\msf{min} 	& = \ln 2, \\
\label{eq:Ns1low2}
\CMcal{S}_0 	& = 2 \frac{\beta}{1+\beta}, \\
\label{eq:Ns1high1}
\CMcal{S}_\infty 	& = 1-e^{-\beta}, \\
\label{eq:Ns1high2}
\CMcal{L}_\infty 	& = \log_{2}\beta - \frac{1}{1-e^{-\beta}} \sum_{k>1} \frac{\beta^k e^{-\beta}}{k!} \log_{2} k .
\end{align}
\smallskip

For the generic case $N_{\msf{s}}>1$, one can show that asymptotics in the wideband regime are the same as above (see eq.~\eqref{eq:Ns1low1}~and~\eqref{eq:Ns1low2}). More precisely, we show in Appendix~\ref{app:dimNsg1wide} that $\E{\lambda}=\beta$ surely for any matrix ensemble where columns of $\bs{S}$ are normalized, and that $\frac{1}{N}\sum_{i=1}^N \lambda_i^2 \xrightarrow{\; p \; } \beta(\beta+1)$. Therefore, from eqs.~\eqref{eq:lowhowto1} and \eqref{eq:lowhowto2}, one has $\eta_\msf{min}=\ln 2$ surely for any $N_\msf{s}$ and $\CMcal{S}_0$ in probability as in eq.~\eqref{eq:Ns1low2}, respectively.

Comparing eqs.~\eqref{eq:Ns1low1}-\eqref{eq:Ns1high2} with DS-CDMA results \cite{VerSha:1999}, shows that TH-CDMA has same wideband asymptotic parameters, $\eta_{\msf{min}}$ and $\CMcal{S}_{0}$, as DS-CDMA, while different high-SNR parameters, $\CMcal{S}_{\infty}$ and $\CMcal{L}_{\infty}$. In particular, in the high-SNR regime, DS-CDMA achieves $\CMcal{S}_{\infty}=\min\{1,\beta\}$ while TH-CDMA achieves $\CMcal{S}_{\infty}=1-e^{-\beta}$, that is, TH-CDMA incurs in a slope penalty given by $e^{-\beta}$. At very high loads, $\beta\gg1$, this penalty becomes negligible, and TH-CDMA high-SNR slope tends to that of DS-CDMA.

Figure~\ref{fig:fig1} shows spectral efficiency $\msf{C}$ (b/s/Hz) of TH-CDMA with $N_\msf{s}=1$ (blue solid line) vs. DS-CDMA (red solid line) as a function of $\Eb/\N$ (dB) with load $\beta=1/2$; simulated data for TH with $N_\msf{s}=1$ are also represented on figure (blue triangles) to highlight agreement with theoretical values. Orthogonal multiple access is also reported for comparison (gray solid line) and represents an upper bound on the sum-rate of a multiuser communication scheme. In the wideband regime, where $\msf{C}\to0$, both TH-CDMA and DS-CDMA achieve $\eta_{\msf{min}}=\ln 2$ and same wideband slope $\CMcal{S}_0$. At the high-SNR regime, where $\Eb/\N\to\infty$, DS achieves larger high-SNR slope than TH. A simulated case of $N_\msf{s}=2$ was also considered in order to understand the effect on $\msf{C}$ of increased $N_\msf{s}$ for TH-CDMA (see blue dots on figure). While for any finite $N_{\msf{s}}$ the spectral efficiency gap between DS-CDMA and TH-CDMA grows as $\Eb/\N$ increases, figure shows that for common values of $\Eb/\N$, \eg $\Eb/\N<20$ dB, $N_\msf{s}=2$ pulses only are sufficient to reduce the gap to very small values. 
Figure~\ref{fig:Cbeta_0} shows spectral efficiency $\msf{C}^\msf{opt}$ (b/s/Hz) for TH with $N_\msf{s}=1$ (blue solid line) and $N_\msf{s}=2$ (dotted line), and for DS (red solid line), for $\Eb/\N=10$ dB. It is shown that TH achieves lower spectral efficiency with respect to DS. However, the loss is negligible for both $\beta\ll 1$ and $\beta\gg 1$. The gap between the two spectral efficiencies can be almost closed with increased, yet finite, $N_\msf{s}$. Simulations suggest that $N_\msf{s}=2$ is sufficient to significantly reduce the gap.

\subsection{Single-User Matched Filter}\label{sub:sumf}
The output of a bank of SUMF is given by eq.~\eqref{eq:discrsynch}, that is, $\bs{y} = \bs{S}\bs{b}+\bs{n}$. Focusing on user $1$, one has:
\begin{align} 
	\msf{y}_1 	& = \bs{s}_1^\t \bs{y} \nonumber \\[-2pt]
			& = \msf{b}_1 + \sum_{k=2}^K \rho_{1k} \msf{b}_k + n_1 \label{eq:sumf2} \\
			& = \msf{b}_{1} + z_{1}, \nonumber 
\end{align}
where $\rho_{1k}\eqdef \bs{s}_1^\t \bs{s}_k$. As shown in \cite{VerSha:1999}, spectral efficiency for binary or spherical DS-CDMA 
when each SUMF is followed by an independent single-user decoder knowing $\bs{S}$ is \cite{VerSha:1999, ShaVer:2001}:
\begin{equation}\label{eq:SUMFDS} C^{\msf{sumf}}_{\textup{DS}}(\beta,\gamma) = \beta \log_{2}{\left(1+\frac{\gamma}{1+\beta\gamma}\right)}\quad \textup{(b/s/Hz)}. \end{equation}
This result is general, and in particular it does not assume that the distribution of neither input nor interference terms is Gaussian. Note, however, that, in this case, Gaussian inputs are optimal. In fact, for long spreading sequences, by virtue of the strong laws of large numbers, one has $\sum_{k=2}^{K}\rho_{1k}^2 \toas  \beta$, and therefore the mutual information per user in bits per channel use is: 
\begin{align}
I(\msf{y}_{1};\msf{b}_{1}|\bs{S})
	& = I\big(\msf{y}_{1};\msf{b}_{1}\big|\rho_{12},\dotsc,\rho_{1K}\big) \nonumber \\
	& = \E{ \log_{2}\!\left( 1+\frac{\gamma}{1 +\gamma \sum_{k=2}^{K}\rho_{1k}^{2} } \right)}   \toas  \log_{2}\!\left( 1+\frac{\gamma}{1 +\gamma\beta}\right).\label{eq:sumf1Sknown}
\end{align}
A similar result does hold for $I(\msf{y}_1;\msf{b}_1)$ as well.

\medskip

When interference is not Gaussian, we may expect spectral efficiency to assume a very different form than above. This will prove to be the case for the mutual information of TH-CDMA assuming Gaussian inputs, with finite $N_\msf{s}$, as investigated below. 

\setlength\ploth{0.375\columnwidth} \setlength\plotw{0.45\columnwidth} 
\begin{figure}[t]
\centering
\includegraphics{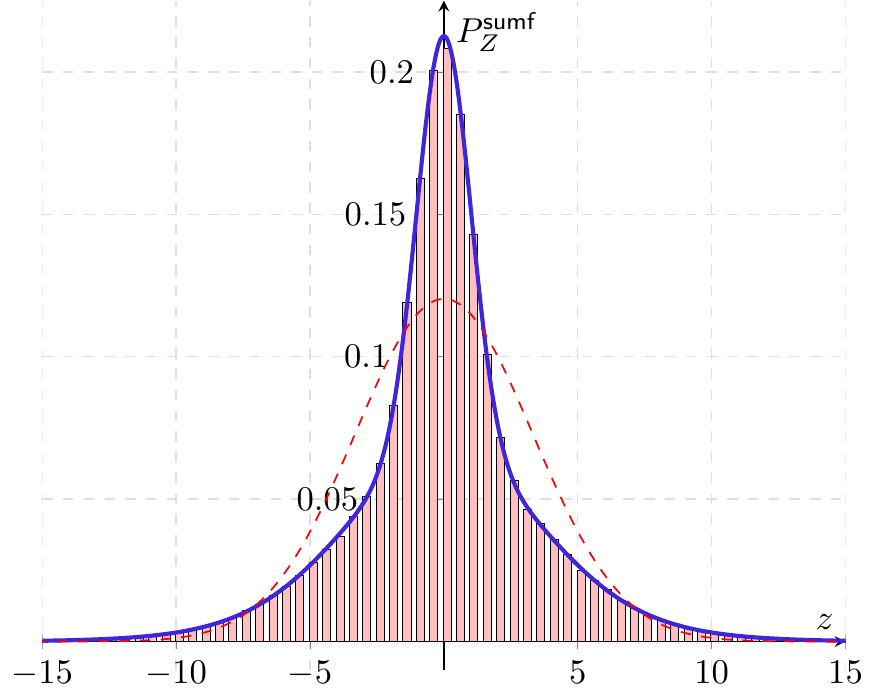}
\caption{Probability density function of the real (or imaginary) component of the noise-plus-interference term of eq.~\eqref{eq:sumf2} for TH sequences with $\gamma=13$ dB, $\beta=1$, and $N_\msf{s}=1$ (blue solid line), and comparison against a Gaussian PDF with same mean and variance (red dashed line). This example shows that, contrary to DS-CDMA, $P_{Z}^\msf{sumf}$ as given in eq.~\eqref{eq:PZsumf} may be, in general, far from Gaussian.}
\label{fig:FIG_SUMF_nonGaussian}
\end{figure}

\smallskip

\begin{theorem}\label{thm:SUMF1} Suppose that $\bs{S}\in\R^{N\times \beta N}$ is a time-hopping matrix with generic $N_\msf{s}<\infty$, and that the receiver is a bank of single-user matched filters followed by independent decoders, each knowing $\bs{S}$. Assuming Gaussian inputs, mutual information $I_{\textup{TH}}^{\msf{sumf}}$ (b/s/Hz) is given by: 
\begin{align} \label{eq:sumfSknownNsleq1}
\hspace{-3mm}I_{\textup{TH}}^{\msf{sumf}}{\negthinspace}(\beta,\gamma,N_\msf{s}) &  \eqdef \beta I(\msf{y}_1;\msf{b}_1|\bs{S}) = \beta\cdot \sum_{k\geq0} \frac{(N_\msf{s}^2\beta)^{k}}{k!}e^{-N_\msf{s}^2\beta} \log_{2}{\negthinspace}\bigg( 1+\frac{\gamma}{1+\frac{k}{N_\msf{s}^2} \gamma} \bigg).
\end{align}
\end{theorem}
\IEEEproof{See Appendix~\ref{app:Nsleq1}.}

In particular, for the $N_\msf{s}=1$ case, mutual information is:
\begin{equation} \label{eq:sumfSknown}
I_{\textup{TH}}^{\msf{sumf}}{\negthinspace}(\beta,\gamma) = \beta \cdot \sum_{k\geq0} \frac{\beta^{k}}{k!}e^{-\beta} \log_{2}{\negthinspace}\bigg( 1+\frac{\gamma}{1+k{\slim} \gamma} \bigg),
\end{equation}
that can be compared to, and interpreted similarly to, eq.~\eqref{eq:sens1}. 

Note that eq.~\eqref{eq:sumfSknown} provides the mutual information of TH-CDMA with $N_{\msf{s}}=1$, and not the spectral efficiency, since Gaussian inputs, rather than optimal ones, are considered. Hence, we know that spectral efficiency will be larger than or equal to $I_{\textup{TH}}^{\msf{sumf}}{\negthinspace}(\beta,\gamma)$. This mutual information expression is, however, sufficient to catch a significant difference between DS-CDMA and TH-CDMA. By comparing eqs.~\eqref{eq:SUMFDS}~and~\eqref{eq:sumfSknown}, we can claim that, while spectral efficiency for DS is bounded at high $\gamma$, being:
\begin{equation}\label{eq:dslimit} \lim_{\gamma\to\infty} C^{\msf{sumf}}_{\textup{DS}}(\beta,\gamma) = \beta \log_{2}\bigg(1+\frac{1}{\beta}\bigg), \end{equation}
spectral efficiency for TH is unbounded. We can indeed derive the below stronger result: 

\begin{corollary} Under the hypotheses of Theorem~\ref{thm:SUMF1}, the high-SNR slope of the mutual information \eqref{eq:sumfSknownNsleq1} of TH is:
\begin{equation}\label{eq:SUMFhslope} \CMcal{S}^{\,\msf{sumf}}_{\infty,\textup{TH}} = \beta e^{-N_\msf{s}^2\beta}.\end{equation}
\end{corollary}

\rem{The maximum slope as a function of $\beta$ is achieved at $\beta=1/N_\msf{s}^2$, for which $\CMcal{S}^{\,\msf{sumf}}_{\infty,\textup{TH}}=1/(eN_\msf{s}^2)$. Since $N_\msf{s}\geq 1$, the global maximum is $1/e$, and the optimum load is $\beta=1$. %
%
%
This behavior directly provides an insight from a design standpoint: at high-SNR, the number of chips such that an increase in $\Eb/\N$ yields a maximum increase in terms of mutual information is equal to the number of users. As a comparison, for optimum decoding, $\CMcal{S}_\infty$ increases monotonically with $\beta$, and its supremum is $\sup \CMcal{S}_\infty=1$.}

Differently from DS, when each user decoder does not have knowledge about cross-correlations of signature sequences of other users, mutual information assumes a very different form, as derived in the following theorem. 

\begin{theorem}\label{theo:SUMF} Suppose that $\bs{S}\in\R^{N\times \beta N}$ is a time-hopping matrix with generic $N_\msf{s}<\infty$, and that the receiver is a bank of single-user matched filters followed by independent decoders, each knowing the signature sequence of the user to decode only. Assuming Gaussian inputs, mutual information $I_{\textup{TH}^{\star}}^{\msf{sumf}}{\negthinspace}(\beta,\gamma,N_\msf{s})$ (bits/s/Hz) is given by: 
\begin{equation}\label{eq:SUMFTH} 
I_{\textup{TH}^{\star}}^{\msf{sumf}}{\negthinspace}(\beta,\gamma,N_\msf{s}) \eqdef  \beta I(\msf{y}_1;\msf{b}_1) = \beta \cdot \,[\, h(P^\msf{sumf}_Y)-h(P^\msf{sumf}_Z)\,],
\end{equation}
where $P^\msf{sumf}_Y$ and $P^\msf{sumf}_Z$ are the two following Poisson-weighted linear combinations of Gaussian distributions:
\begin{align}
	P^\msf{sumf}_Y & = \sum_{k\geq 0} \frac{(\beta N_\msf{s}^2)^k}{k!} e^{-\beta N_\msf{s}^2} \CNorm{0}{1+\gamma+k\gamma/N_\msf{s}^2}, \label{eq:PYsumf}\\
	P^\msf{sumf}_Z & = \sum_{k\geq 0} \frac{(\beta N_\msf{s}^2)^k}{k!} e^{-\beta N_\msf{s}^2} \CNorm{0}{1+k\gamma/N_\msf{s}^2}\label{eq:PZsumf}.
\end{align}
\end{theorem}
\IEEEproof{See Appendix~\ref{app:SUMFproof}.}

Despite decoders partial knowledge of $\bs{S}$, a same high-SNR slope as that achieved when decoders have knowledge of $\bs{S}$ is verified in the $N_\msf{s}=1$ case: 
\begin{corollary} Under the hypotheses of Theorem~\ref{theo:SUMF}, the high-SNR slope of the mutual information $I_{\textup{TH}^{\star}}^{\msf{sumf}}{\negthinspace}(\beta,\gamma,1)$ is:
\begin{equation}\label{eq:highSNRslopeSUMFTH}
\qquad\CMcal{S}^{\,\msf{sumf}}_{\infty,\textup{TH}^{\star}} = \beta e^{-\beta}. 
\end{equation}
\end{corollary}

\setlength\ploth{0.6\columnwidth} \setlength\plotw{0.85\columnwidth} 
\begin{figure*}[t]
\centering
\includegraphics{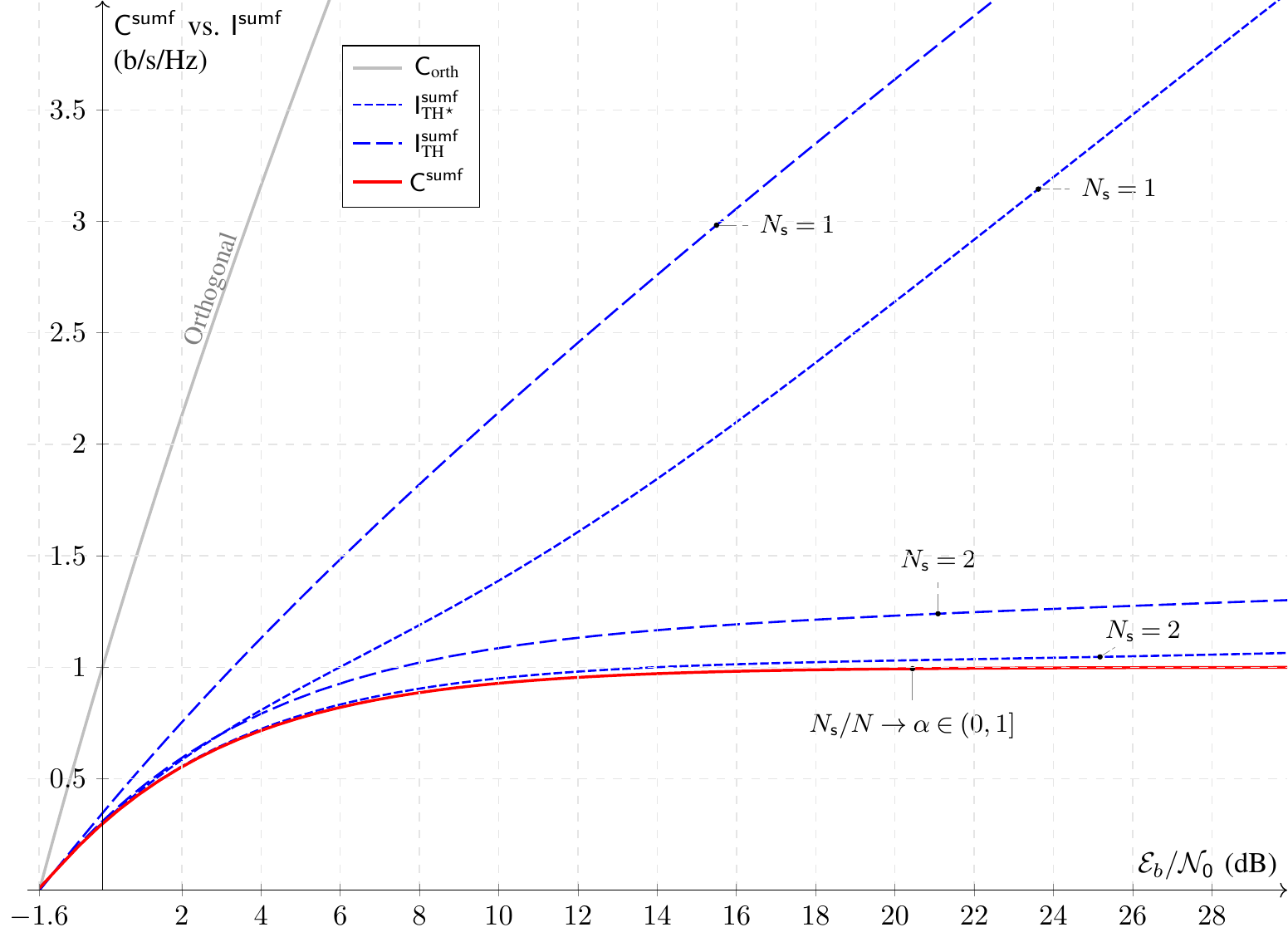}
\caption{Spectral efficiency $\msf{C}^{\msf{sumf}}$ vs. $\msf{I}^{\msf{sumf}}$ (b/s/Hz) as a function of $\Eb/\N$ (dB) with load $\beta=1$. Closed form expressions of spectral efficiency vs. mutual information are plotted in solid vs. dashed lines. Simulated mutual information is represented by dotted lines. On figure: SUMF, TH-CDMA, $N_\msf{s}=1$, $N_\msf{s}=2$ and $N_\msf{s}=5$, blue dashed lines; SUMF, TH-CDMA$^\star$, $N_\msf{s}=1$, blue dashed line; DS-CDMA, red solid line; TH-CDMA with $N_\msf{s}=\alpha N$ when $N\to\infty$, blue solid line, coinciding with red solid line; TH-CDMA$^{\star}$ with $N_{\msf{s}}=2$ and $N_{\msf{s}}=5$, blue dotted lines. Note on figure the crossover of SUMF, TH-CDMA, $N_\msf{s}=2$ and SUMF, TH-CDMA$^\star$, $N_\msf{s}=2$, that shows an example of mutual information becoming greater than conditional mutual information. For reference, orthogonal multiple-access in gray line.}
\label{fig:C_Eb_1}
\end{figure*}

Based on eq.~\eqref{eq:PZsumf}, it can be easily checked that the kurtosis of the interference-plus-noise $z_1$, that we denote $Z$ since it is independent of the user, is:
\begin{equation} \label{eq:z1kurtosis}
\kappa_{Z} \eqdef \frac{\E{|Z|^{4}}}{\E{|Z|^{2}}^{2}} = 2 + \frac{2}{N_\msf{s}^2}\cdot \frac{\beta \gamma^{2}}{(1+\beta\gamma)^{2}}, 
\end{equation}
that is always greater than $2$, hence showing non-Gaussianity of $Z$ for any $\beta$, $\gamma$ and $N_\msf{s}$. This non-Gaussian nature is represented on Fig.~\ref{fig:FIG_SUMF_nonGaussian}, that shows the interference-plus-noise PDF $P^{\msf{sumf}}_Z$ (solid blue line on figure), as given by eq.~\eqref{eq:PZsumf} when $\beta=1$, $\gamma=13$ dB and $N_\msf{s}=1$, vs. a Gaussian distribution with same mean and variance (red dashed line on figure). As shown by figure, $P^{\msf{sumf}}_Z$, that is a linear combination, or ``mixture,'' of Gaussian distributions with Poisson weights, cannot be reasonably approximated with a single Gaussian distribution; hence, the Standard Gaussian Approximation does not hold in general. This is the reason for the mutual information gap between DS and TH. 

\medskip

The wideband regime is not affected by decoders' knowledge about crosscorrelations between signature sequences, as summarized by the below corollary, which proof is omitted for brevity. 
\begin{corollary}
The wideband regime parameters derived from either eq.~\eqref{eq:sumfSknown} or eq.~\eqref{eq:SUMFTH} are $\eta_{\msf{min}}=\ln 2$ and:
\begin{equation} \label{eq:s0thsumfgivenS}
\CMcal{S}_{0,\textup{TH}}^{\msf{sumf}}=\frac{2\beta}{1+2\beta}.
\end{equation}
\end{corollary}

\medskip

%
%
%
%
%
%
%
%
Differently from above, where $N_\msf{s}$ is finite and does not depend on $N$, we now investigate the case $N_{\msf{s}}=\alpha N$ with $\alpha\in(0,1)$, while $N\to\infty$. We show, using an approach similar to that developed in \cite{VerSha:1999}, that spectral efficiency of a TH channel with $N_{\msf{s}}=\alpha N$, $\alpha\in(0,1)$, is equal to that of a DS system, irrespectively of $\alpha\in(0,1)$. 

\begin{theorem}\label{theo:SUMFcapacity} Suppose that $\bs{S}\in\R^{N\times \beta N}$ is a time-hopping matrix with $N_\msf{s}=\alpha N$, $\alpha\in(0,1)$, and that the receiver is a bank of single-user matched filters followed by independent decoders knowing cross-correlations and input distributions of interfering users. Capacity $C^{\msf{sumf}}(\rho_{12},\dotsc,\rho_{1K},P_{\msf{b}_{2}},\dotsc,P_{\msf{b}_{K}})$ of the single-user channel of eq.~\eqref{eq:sumf2}, expressed in bits per user per channel use, converges almost surely to:
\begin{equation}\label{eq:specSUMFgen} C^{\msf{sumf}}(\rho_{12},\dotsc,\rho_{1K},P_{\msf{b}_{2}},\dotsc,P_{\msf{b}_{K}})  \toas  \log_{2}\bigg(1+\frac{\gamma}{1+\beta\gamma}\bigg),\end{equation}
irrespective of $\alpha$. 
\end{theorem}
\IEEEproof{See Appendix~\ref{app:sumfcapacity}.}

\smallskip

Based on eq.~\eqref{eq:specSUMFgen}, spectral efficiency coincides with that of DS sequences, as given by eq.~\eqref{eq:SUMFDS}. As a matter of fact, Theorem~\ref{theo:SUMFcapacity} is the generalization of a result of Verd\'u and Shamai \cite{VerSha:1999} to TH matrices where the fraction of nonzero entries is $\alpha$, to which it reduces for $\alpha=1$.

\begin{figure*}[t]
\centering
\setlength\ploth{0.375\columnwidth} \setlength\plotw{0.45\columnwidth}
\includegraphics{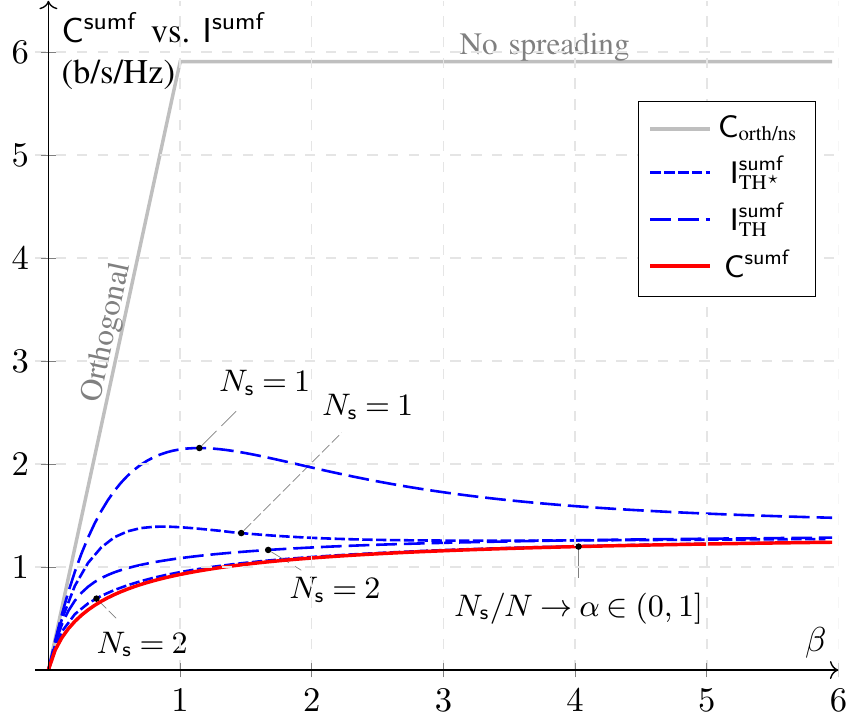}
\caption{Spectral efficiency $\msf{C}^\msf{sumf}$ vs. mutual information $\msf{I}^\msf{sumf}$ as a function of $\beta$ for fixed $\Eb/\N=10$ dB. DS and TH with $N_\msf{s}/N\to\alpha\in(0,1)$ are shown in red solid line, and represent the worst performance on figure. Dashed lines correponds to either TH knowing $\bs{S}$ (large dashing) or TH where decoders know the spreading sequence of the user to decode only (small dashing). Orthogonal access is reported for reference (gray solid line).}
\label{fig:Cbeta_1}
\end{figure*}

Figure~\ref{fig:C_Eb_1} shows spectral efficiency $\msf{C}^\msf{sumf}$ vs. mutual information $\msf{I}^\msf{sumf}$ (b/s/Hz) as a function of $\Eb/\N$ (dB) for DS-CDMA (eq.~\eqref{eq:SUMFDS}, red solid line on figure), TH-CDMA knowning cross-correlations between users (eq.~\eqref{eq:sumfSknown}, blue large-dashed lines) and TH-CDMA without knowing cross-correlations between users, indicated as TH-CDMA$^{\star}$ (eq.~\eqref{eq:SUMFTH}, blue small-dashed line), with unit load $\beta=1$. Spectral efficiency of TH-CDMA when $N_\msf{s}=\alpha N$, $\alpha\in(0,1)$, as $N\to\infty$, is equal to that of DS (c.f. eq.~\eqref{eq:specSUMFgen}, red solid line). As previously, the orthogonal case (gray solid line) is shown for reference. Note that spectral efficiency is bounded in DS-CDMA and in TH-CDMA when $N_\msf{s}=\alpha N$, $\alpha\in(0,1)$, as $N\to\infty$; the value of the limit is $1$ on figure (c.f. eq.~\eqref{eq:dslimit}). On the contrary, mutual information is not bounded for both TH-CDMA and TH-CDMA$^\star$; in particular, when $N_\msf{s}=1$, both TH-CDMA and TH-CDMA$^{\star}$ grow with similar slope as $\Eb/\N$ increases. 
Mutual information of systems using multiple pulses per symbol is shown for TH-CDMA$^\star$ with $N_{\msf{s}}=2$ (small-dashed line) and for TH-CDMA with $N_\msf{s}=2$ (eq.~\eqref{eq:sumfSknownNsleq1}, large-dashed line). These $N_\msf{s}\neq 1$ cases show that mutual information decreases with respect to the one pulse per symbol case. 
%
%
Figure~\ref{fig:Cbeta_1} shows spectral efficiency $\msf{C}^\msf{sumf}$ (b/s/Hz) as a function of $\beta$ for fixed $\Eb/\N=10$ dB. Similarly as on fig.~\ref{fig:C_Eb_1}, TH with $N_\msf{s}=1$ outperforms other schemes, with and without complete knowledge of $\bs{S}$. As $\beta\to\infty$, interference becomes increasingly Gaussian, and mutual information of TH reduces to that of DS, tending to the same limit $1/\ln 2$. 

\subsection{Decorrelator and MMSE}\label{sub:deco}

The output of a bank of decorrelators, following the discrete channel $\bs{y} = \bs{S}\bs{b}+\bs{n}$ (c.f. eq.~\eqref{eq:discrsynch}), is given by:
\begin{equation}\label{eq:deco} \bs{r} = \bs{S}^+ \bs{y} = \bs{S}^+\bs{S}\bs{b}+\bs{S}^+\bs{n}, \end{equation}
where $\bs{S}^+$ denotes the Moore-Penrose pseudoinverse; if $\bs{R}=\bs{S}^\t\bs{S}$ is invertible, then $\bs{S}^+=(\bs{S}^\t\bs{S})^{-1}\bs{S}^\t$, otherwise $\bs{S}^+$, according to the Tikhonov regularization, exists and can be computed as the limit $(\bs{S}^\t\bs{S}+\alpha \bs{I})^{-1}\bs{S}^\t$ as $\alpha\to0^+$. 

In DS-CDMA, for any fixed $\beta\in(0,1)$, $\bs{S}$ is almost surely full rank as $N\to\infty$, and therefore, $\bs{R}$ is almost surely invertible, in which case eq.~\eqref{eq:deco} becomes:
\begin{equation}\label{eq:decoinv} \bs{r} = \bs{b} + \bs{z},\end{equation}
where $\bs{z}\sim\CNorm{\bs{0}}{\bs{R}^{-1}\N}$. Assuming independent single-user decoders, spectral efficiency is \cite{VerSha:1999}: 
\begin{equation}\label{eq:DSdeco}
C^{\msf{deco}}_{\textup{DS}}(\beta,\gamma) = \beta \log_{2}(1+\gamma(1-\beta)).
\end{equation}

The output of a bank of MMSE filters observing $\bs{y} = \bs{S}\bs{b}+\bs{n}$ (c.f. eq.~\eqref{eq:discrsynch}) is:
\begin{align} 
\bs{r} 	& = \bs{W}^{\t} \bs{y} = \bs{W}^{\t}\bs{S}\bs{b}+\bs{W}^{\t} \bs{n}  \nonumber \\
		& = \bs{G}\bs{b}+\bs{\nu} \label{eq:mmse2},
\end{align}
where $\bs{W}^\t$ is defined as follows:
\begin{equation} \label{eq:wtmmse}
\bs{W}^{\t} \eqdef \bigg(\!{\bs{S}^\t\bs{S}+\displaystyle{\frac{1}{\gamma}}\bs{I}}\bigg)^{\!\!-1}\bs{S}^\t = \bs{S}^\t\,\bigg(\bs{S}\bs{S}^\t+\displaystyle{\frac{1}{\gamma}}\bs{I}\bigg)^{\!-1}.
\end{equation}
Note that, as well known, MMSE and decorrelator coincide as $\gamma\to\infty$. In DS-CDMA, for any fixed $\beta>0$, it was shown in \cite{VerSha:1999} that:
\begin{equation} \label{eq:mmseds}
C_\textup{DS}^{\msf{mmse}}(\beta,\gamma)=\beta\log_{2}\bigg(1+\gamma-\frac{1}{4}\CMcal{F}(\beta,\gamma)\bigg),
\end{equation}
where:
\[ \CMcal{F}(\beta,\gamma) = \Big[ \sqrt{1+\gamma \ell^+}-\sqrt{1+\gamma\ell^-} \,\Big]^{2}, \]
being $\ell^{\pm}=(1\pm\sqrt{\beta})^2$ as in eq.~\eqref{eq:mpdensity}.

We can treat both decorrelator and MMSE as special cases of the linear operator
\[ \bs{W}^\t(\alpha)\eqdef (\bs{S}^\t \bs{S}+\alpha \bs{I})^{-1} \bs{S}^\t = \bs{S}^\t(\bs{S}\bs{S}^\t+\alpha \bs{I})^{-1}, \]
for $\alpha\to0^+$ and $\alpha=1/\gamma$, respectively. Similarly as eq.~\eqref{eq:mmse2}, one has:
\begin{equation}\label{eq:genlin} 
 \bs{r}(\alpha)=\bs{W}^\t(\alpha)\bs{y}=\bs{G}(\alpha) \bs{b}+\bs{\nu}(\alpha), 
 \end{equation}
where dependence on $\alpha$ is now made explicit, and the output for user $1$ is:
\begin{equation}\label{eq:genlinuser1a} 
r_1 = G_{11} \msf{b}_1 + \sum_{k=2}^K G_{1k} \msf{b}_k + \nu_1. 
\end{equation}
For $N_\msf{s}=1$, a closed form expression for the generic element of $\bs{G}(\alpha)$ is derived in Appendix~\ref{app:GENLINproof}, and reads as:
\begin{equation}\label{eq:genlin2}  G_{ij}(\alpha) = \rho_{ij}\cdot \frac{1}{\alpha+v_i}, \end{equation}
where $v_i$ is:
\begin{equation}\label{eq:genlin3}  v_{i} = \sum_{k=1}^K \indicator\{\rho_{ik}\neq 0\} = \sum_{k=1}^{K} |\rho_{ik}| = \sum_{k=1}^{K} |\rho_{ki}|. \end{equation}
Denote with $\Jidx_{j}$ the following set: 
$\Jidx_{j} \eqdef \{ k\in\intint{1}{K}\colon \rho_{jk}\neq 0 \}$. %
Hence, $v_{j}=|\Jidx_{j}|$ is the cardinality of $\Jidx_{j}$. Denote with $\Jidx_{j}'\eqdef \Jidx_{j}\backslash\{j\}$. Since $j\in\Jidx_{j}$, one has $v_j'\eqdef |\Jidx_j'|=v_j-1$. %
We can rewrite eq. \eqref{eq:genlinuser1a} as follows:
\begin{align} \label{eq:genlinuser1givenn}
	r_1 		& = \frac{1}{\alpha+ v_1} \msf{b}_1 + \frac{1}{\alpha+ v_1} \sum_{k\in\Jidx_{1}'} \rho_{1k} \msf{b}_k + \nu_1. 
\end{align}
Note that $\rho_{1k} \msf{b}_k$ for $\rho_{1k}\neq 0$ is distributed as $\msf{b}_{k}$, and $\nu_{1}$ given $v_1$ is complex Gaussian with zero mean and conditional variance:
\[ \Var{\nu_{1} \big| \,v_1} = \N\cdot \frac{1}{(\alpha+ v_1)^{2}}. \]
Since the distribution of both $r_1$ conditioned on $\bs{S}$ and $r_1$ conditioned on $\msf{b}_1$ and $\bs{S}$ is complex Gaussian, $I(\msf{b}_{1};r_{1}|\bs{S}) $ expressed in bits per user per channel use is: 
\begin{align*} I(\msf{b}_{1};r_{1}|\bs{S})  = I(\msf{b}_{1};r_{1}|v_1')
							& = \Expectation\Bigg[{ \, \log_{2}\!\left( 1+\frac{ \En/(\alpha+ v_1')^{2} }{ (v_1'\En+\N)/(\alpha+ v_1')^{2} } \right) \,}\Bigg] \\
							& = \Expectation\Bigg[{ \, \log_{2}\!\left( 1+\frac{ \gamma }{ v_1'\gamma+1 } \right) \,}\Bigg],  
\end{align*} 
Since $v_1'\sim \textup{Binomial}(K-1,1/N)$, in the LSL one has $v_1'\xrightarrow{d}\Pois{\beta}$. Therefore, we proved the following:
\begin{theorem}\label{theo:genlin} %
Suppose that $\bs{S}\in\R^{N\times \beta N}$ is a time-hopping matrix with $N_\msf{s}=1$, and that the receiver is a bank of either decorrelators $(\alpha=0)$ or MMSE filters $(\alpha=1/\gamma)$ followed by independent decoders, each knowing $\bs{S}$. Assuming Gaussian inputs, mutual information $I_{\textup{TH}}^{\alpha}$ (b/s/Hz) is given by:
\begin{align}\label{eq:Igenlin}
I_{\textup{TH}}^{\alpha}(\beta,\gamma)\eqdef \beta I(\msf{b}_{1};r_{1}|\bs{S}) = \beta\sum_{k\geq 0} \frac{\beta^{k}e^{-\beta}}{k!} \log_{2}\!\bigg( 1+\frac{\gamma}{k\gamma+1} \bigg).
\end{align}
\end{theorem}
Since eq.~\eqref{eq:Igenlin} does not depend on $\alpha$ and is equal to eq.~\eqref{eq:sumfSknown} for SUMF, one explicitly has ${I}_\textup{TH}^\alpha={I}_\textup{TH}^\msf{sumf}={I}_\textup{TH}^\msf{mmse}={I}_\textup{TH}^\msf{deco}$. 
With minor modifications of the above argument, it is possible to show that a similar result does hold for any linear receiver $\bs{W}^\t(\alpha)$, $\alpha>0$, under the assumption $N_\msf{s}=1$. Therefore, results for SUMF can be extended verbatim to both decorrelator and MMSE receivers, when $N_\msf{s}=1$. This result suggests a striking difference with respect to DS, where spectral efficiency depends on the adopted linear receiver: In TH with $N_\msf{s}=1$, SUMF, decorrelator and MMSE all result in the same mutual information. 

\setlength\ploth{0.5\columnwidth} \setlength\plotw{0.725\columnwidth} 
\begin{figure}[t]
\centering
\includegraphics{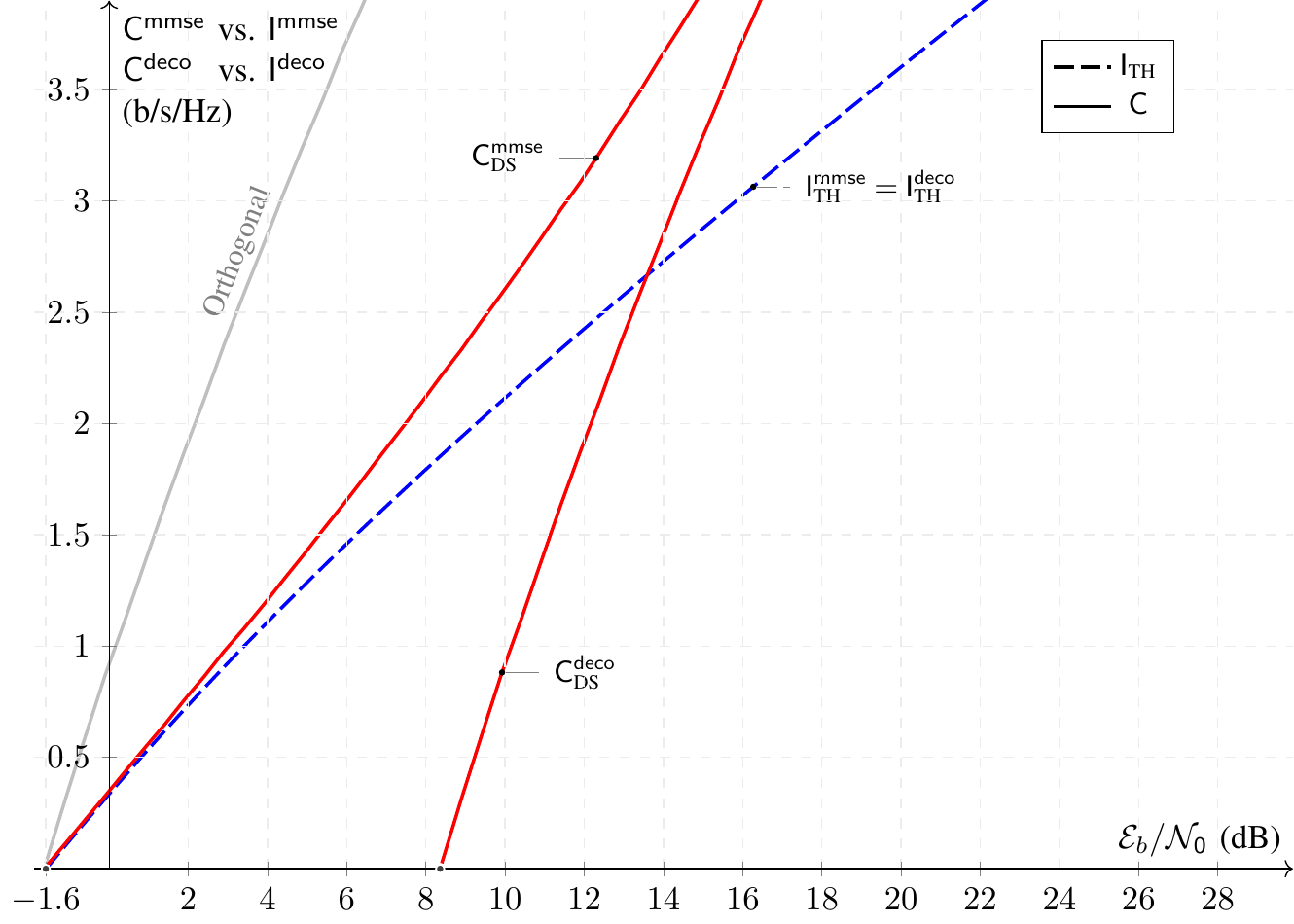}
\caption{Spectral efficiency $\msf{C}^{\msf{mmse}}$ vs. mutual information $\msf{I}^{\msf{mmse}}$ and $\msf{C}^{\msf{deco}}$ vs. $\msf{I}^{\msf{deco}}$ (b/s/Hz) as a function of $\Eb/\N$ (dB) with load $\beta=0.9$. Mutual information of TH-CDMA with $N_\msf{s}=1$ (blue dashed line) vs. spectral efficiency of DS-CDMA (red solid lines), for decorrelator and MMSE receivers, is shown. It is also shown orthogonal access (gray line) for reference.}
\label{fig:C_Eb_2}
\end{figure}

In order to compare DS and TH for decorrelator and MMSE, we separate the analysis for systems with $\beta<1$ and $\beta>1$, referred to as underloaded and overloaded systems, respectively.

\smallskip\noindent\textit{Underloaded system} $(\beta<1)$. 

Decorrelation in DS allows to achieve the maximum high-SNR slope, $\CMcal{S}^\msf{deco}_{\infty,\textup{DS}}=\beta$, that is equal to that of orthogonal multiple access. On the contrary, TH does not fully exploit the capabilities of CDMA in the high-SNR regime, since $\CMcal{S}^\msf{deco}_{\infty,\textup{TH}}=\CMcal{S}^{\,\msf{sumf}}_{\infty,\textup{TH}}=\beta e^{-\beta}\leq \beta$. This behavior follows directly from cross-correlation properties of signature sequences of DS vs. TH: In DS, the almost sure linear independence of signature sequences, that holds for any $\beta\in(0,1)$, makes $\bs{R}=\bs{S}^\t\bs{S}$ almost sure invertible, and thus interference can be mostly removed, which is not the case of TH (c.f. Fig.~\ref{fig:figrank} and Theorem~\ref{thm:rank1}). However, the optimal high-SNR slope in DS comes at the expense of a  minimum $\Eb/\N$ equal to $(\ln 2)/(1-\beta)$, that can be much larger than that achieved by TH, namely $\ln 2$; in particular, as $\beta\to1^{-}$, the minimum energy-per-bit for DS with decorrelator grows without bound. Therefore, decorrelation with DS should to be considered in a very low load, high-SNR regime only: in this region, it outperforms TH. It can be shown, by comparing eqs.~\eqref{eq:mmseds} and \eqref{eq:DSdeco}, that in DS spectral efficiency of MMSE is always larger than that of decorrelator. In particular, it achieves a minimum energy-per-bit equal to $\ln 2$, which is optimal, and also an optimal high-SNR slope. 

\smallskip\noindent\textit{Overloaded system} $(\beta>1)$. 

 Spectral efficiency of TH and DS with MMSE is similar in the low-SNR regime, with same minimum energy-per-bit and wideband slope. At high-SNR, mutual information of TH is unbounded, while spectral efficiency of DS is bounded, as in the SUMF case. In particular, while the high-SNR slope of TH is equal to $\CMcal{S}_{\infty,\textup{TH}}^\msf{sumf}(\beta)=\beta e^{-\beta}$ for any $\beta$, the high-SNR slope of DS with MMSE is:
\[ \CMcal{S}_{\infty,\textup{DS}}^\msf{mmse}(\beta)=\beta \;\indicator\{\beta\in[0,1)\}+\frac{1}{2} \indicator\{\beta=1\}+0\cdot\indicator\{\beta>1\}, \]
which implies that, as $\Eb/\N\to\infty$, $\msf{C}_\textup{DS}^\msf{mmse}$ is infinite for $\beta\leq 1$, while it is finite for $\beta>1$, and equal to (c.f. eq.~\eqref{eq:mmseds}) \cite{VerSha:1999}:
\begin{equation}\label{eq:dsmmselim} \lim_{\gamma\to\infty} C_\textup{DS}^{\msf{mmse}}(\beta,\gamma)=\beta \log_2\frac{\beta}{\beta-1}. \end{equation}
By comparing this result with eq.~\eqref{eq:dslimit}, that refers to SUMF, one also notes that the two limits are different, although as $\beta\to\infty$ both tend to $1/\ln 2$. 

\begin{figure*}[t]
\centering
\setlength\ploth{0.375\columnwidth} \setlength\plotw{0.45\columnwidth}
\includegraphics{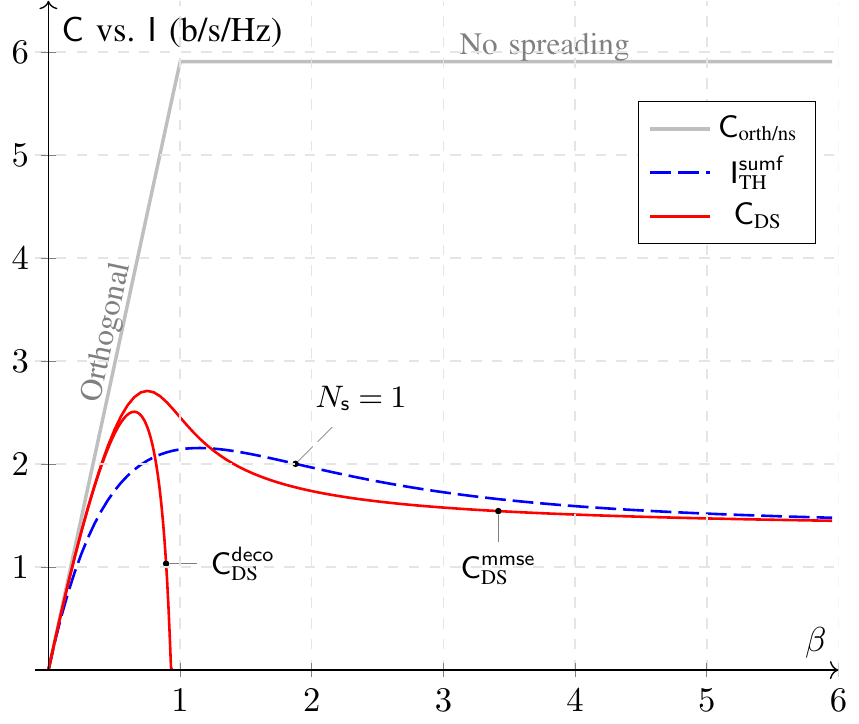}
\caption{Spectral efficiency $\msf{C}_\textup{DS}^\msf{mmse}$ and $\msf{C}_\textup{DS}^\msf{deco}$ vs. mutual information $\msf{I}^\msf{sumf}_\textup{TH}$ (b/s/Hz) as a function of $\beta$ for DS-CDMA (red solid lines) and TH-CDMA (blue dashed line), when $\Eb/\N=10$ dB. Orthogonal access is reported for reference (gray solid line).}
\label{fig:Cbeta_2}
\end{figure*}

\begin{figure*}[t]
\centering
\setlength\ploth{0.375\columnwidth} \setlength\plotw{0.45\columnwidth}
\includegraphics{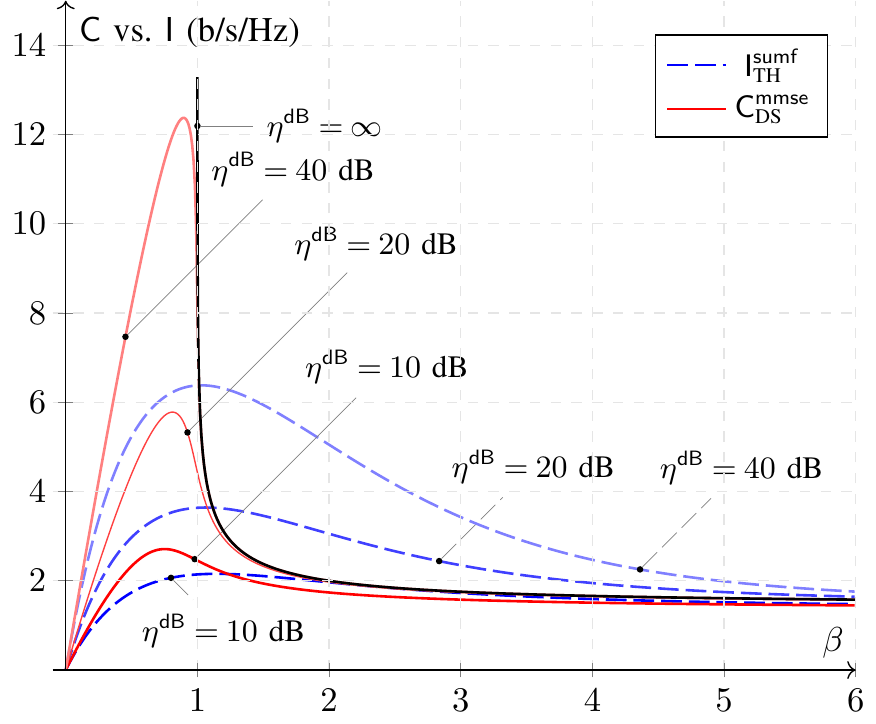}
\caption{Spectral efficiency $\msf{C}_\textup{DS}^\msf{mmse}$ vs. mutual information $\msf{I}^\msf{sumf}_\textup{TH}$ (b/s/Hz) as a function of $\beta$ for DS-CDMA (red solid lines) and TH-CDMA (blue dashed lines), for values of $\eta^\msf{dB}:=\Eb/\N\in\{10,30,50\}$ dB. Asymptotic value of  $\msf{C}_\textup{DS}^\msf{mmse}$ for $\eta^\msf{dB}\to\infty$ is also shown for reference (thin solid black line).}
\label{fig:Cbeta_2b}
\end{figure*}

Figure~\ref{fig:C_Eb_2} shows spectral efficiency $\msf{C}^\msf{mmse}$ and $\msf{C}^\msf{deco}$ vs. mutual information $\msf{I}^\msf{mmse}$ and $\msf{I}^\msf{deco}$ (b/s/Hz) as a function of $\Eb/\N$ (dB) for DS (red solid lines) and TH (blue dashed line), when $\beta=0.9$. Orthogonal access is also shown for reference (gray solid line). The choice of $\beta=0.9$ represents a scenario with high interference where eq.~\eqref{eq:DSdeco} is still valid, and DS with decorrelation still comparable. MMSE and decorrelator receivers achieve a same mutual information for TH: in the low-SNR regime, $\msf{I}^\msf{mmse}_\textup{TH}=\msf{I}^\msf{deco}_\textup{TH}$ and $\msf{C}_\textup{DS}^\msf{mmse}$ have similar behavior, that departs as $\Eb/\N$ increases. Decorrelator with DS achieves the maximum high-SNR slope, which is equal to that of the orthogonal access: note that the two curves on figure are, in fact, translated. This is not the case for TH, for $\bs{S}$ is not full rank with high probability, and the high-SNR slope is indeed lower. It is shown on figure that DS with MMSE outperforms linear receivers with TH: this is due to the particular choice of $\beta$. Figure~\ref{fig:Cbeta_2} shows spectral efficiency $\msf{C}^\msf{mmse}_\textup{DS}$ and $\msf{C}^\msf{deco}_\textup{DS}$ (red solid lines) vs. mutual information $\msf{I}^\msf{sumf}_\textup{TH}=\msf{I}^\msf{deco}_\textup{TH}=\msf{I}^\msf{mmse}_\textup{TH}$ (blue dashed line) as a function of $\beta$, when $\Eb/\N=10$ dB. This figure shows that MMSE with DS is outperformed by TH for large $\beta$: in particular, there exists a minimum value of $\beta$, say $\bar\beta$, in general depending on $\Eb/\N$, beyond which the mutual information of TH is higher than the spectral efficiency of DS, although both tending to a same limit as $\beta\to\infty$, that is, $1/\ln 2$. While it is difficult to study $\bar\beta$ as a function of $\Eb/\N$, the above discussion on the high-SNR slope of DS suggest that $\beta=1$ marks a transition in DS behavior as $\Eb/\N\to\infty$. Figure~\ref{fig:Cbeta_2b} shows $\msf{C}^\msf{mmse}_\textup{DS}$ (red solid lines) and $\msf{I}^\msf{sumf}_\textup{TH}=\msf{I}^\msf{deco}_\textup{TH}=\msf{I}^\msf{mmse}_\textup{TH}$ (blue dashed line) as a function of $\beta$, for different values of $\eta^\msf{dB}=10\log_{10} (\Eb/\N)$. Figure shows that, as $\eta^\msf{dB}$ increases, spectral efficiency of DS grows linearly for $\beta\ll 1$, and at about $\beta=1$ quickly drops towards the limit value given by eq.~\eqref{eq:dsmmselim}, while spectral efficiency of TH remains smooth for any load in the neighborhood of $\beta=1$ and increases monotonically with $\eta^\msf{dB}$.


\setlength\ploth{0.65\columnwidth} \setlength\plotw{0.95\columnwidth} 
\begin{figure*}[t]
\centering
\includegraphics{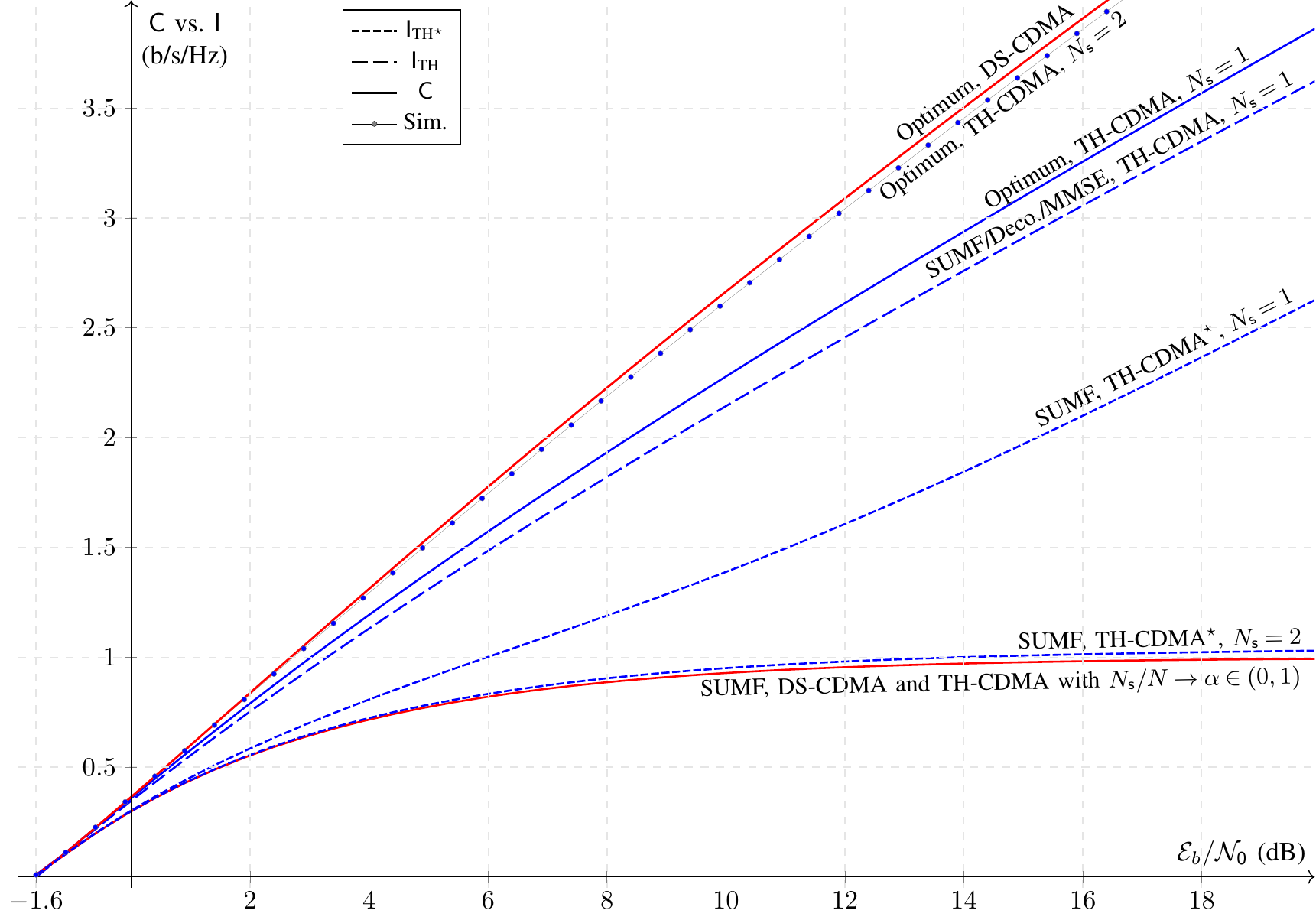}
\caption{Spectral efficiency $\msf{C}$ vs. mutual information $\msf{I}$ (b/s/Hz) as a function of $\Eb/\N$ (dB) with load $\beta=1$. Optimum vs. linear receivers are shown. Top curve shows $\msf{C}^\msf{opt}_\textup{DS}$ for optimum decoding in DS-CDMA (red solid line). Bottom curve shows $\msf{C}$ for SUMF, DS-CDMA (red solid line) coinciding with TH-CDMA when $N_\msf{s}$ goes to infinity proportionally to $N$, \ie, $\lim_{N\to\infty}N_\msf{s}/N=\alpha\in(0,1)$ (red solid line). In between these two extremes: $\msf{C}_\textup{TH}^\msf{opt}$ curve for optimum decoding, TH-CDMA, $N_\msf{s}=2$, simulated values (dotted blue line); $\msf{C}_\textup{TH}^\msf{opt}$ curve for optimum TH-CDMA, $N_{\msf{s}}=1$ (blue solid line); $\msf{I}_\textup{TH}$ curve for linear receivers, TH-CDMA, $N_\msf{s}=1$ (blue large-dashed line); $\msf{I}_{\textup{TH}^\star}^\msf{sumf}$ curve for SUMF, TH-CDMA$^\star$, $N_\msf{s}=1$ and $N_\msf{s}=2$ (blue small-dashed line). 
}
\label{fig:C_Eb_synopsis}
\end{figure*}

\subsection{Synopsis of the TH-CDMA case}\label{sub:synopsis}

Figure~\ref{fig:C_Eb_synopsis} shows spectral efficiency $\msf{C}$ or mutual information $\msf{I}$ (b/s/Hz) vs. $\Eb/\N$ (dB) for the two extreme cases of optimum decoding and SUMF receivers, when $\beta=1$. Curves derived from closed form expressions of spectral efficiency are shown for optimum decoding when $N_\msf{s}=1$ (top blue solid line), and SUMF when $N_\msf{s}=\alpha N$ as $N\to\infty$ and $\alpha\in(0,1)$ (bottom blue solid line). Curves derived from closed form expressions of mutual information assuming Gaussian inputs are shown for SUMF, TH-CDMA (blue dashed line, see label on figure) and SUMF, TH-CDMA$^\star$ (blue dashed line, see label on figure), when $N_\msf{s}=1$ and $N_\msf{s}=2$. Finding closed form expressions of spectral efficiency of optimum decoding with generic $N_\msf{s}>1$ finite remains an open problem. Simulations provide, however, insights into the behavior of spectral efficiency for this particular case, as shown by $\msf{C}^\msf{opt}_\textup{TH}$ with $N_\msf{s}=2$ (blue dotted line). TH behavior is delimited by DS curves, with optimum decoding vs. SUMF (top and bottom red lines). Both upper and lower curves are approached by TH as $N_\msf{s}$ increases; in particular, we showed that the lower curve describes, in fact, TH when $N_\msf{s}=\alpha N$, $\alpha\in(0,1)$, as $N\to\infty$. In between these two extremes lie TH curves with optimum vs. linear receivers. In particular, for $N_\msf{s}=1$ (maximum energy concentration), mutual information of a receiver as simple as SUMF is not bounded, and also close to optimum decoding with $N_{\msf{s}}=1$. Furthermore, a lack of knowledge in cross-correlations of spreading codes provokes a drop of performance that is, however, not sufficient to degrade mutual information to DS spectral efficiency, with any finite $N_\msf{s}$.

\begin{figure*}[t]
\centering
\setlength\ploth{0.375\columnwidth} \setlength\plotw{0.45\columnwidth}
\includegraphics{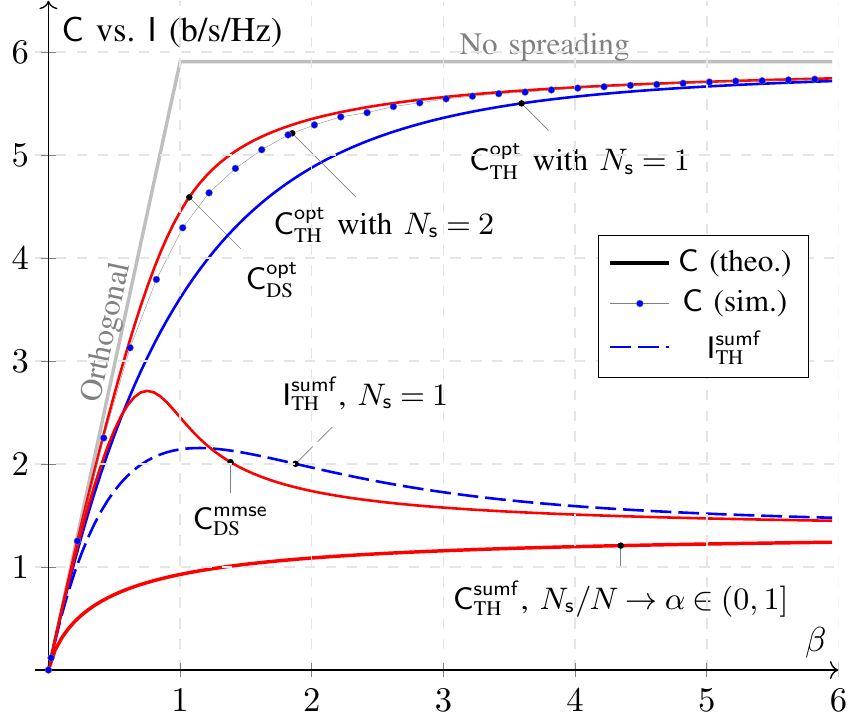}
\caption{Spectral efficiency $\msf{C}$ or mutual information $\msf{I}$ (b/s/Hz) as a function of $\beta$ for DS and TH, when $\Eb/\N=10$ dB, with optimum and SUMF receivers. Orthogonal access and DS with MMSE receiver are reported for reference.}
\label{fig:Cbeta_3}
\end{figure*}

Figure~\ref{fig:Cbeta_3} compares either spectral efficiency $\msf{C}$ or mutual information $\msf{I}$ (b/s/Hz), as a function of $\beta$, for DS and TH, when $\Eb/\N=10$ dB. Both DS and TH have similar behaviors when $\beta\ll 1$, for linear and optimum receivers. Irrespective of $\beta$, spectral efficiency of DS with optimum decoding is larger than that achieved by TH, the gap being almost closed when $N_\msf{s}>1$ finite. Conversely, among linear receivers and access schemes, it is shown that DS with SUMF has the lowest spectral efficiency, which is equal to that of TH when the number of pulses is asymptotically a nonzero fraction of the number of chips. The largest spectral efficiency in DS is obtained with MMSE, which is greater than the mutual information of TH when load is lower than a threshold $\bar{\beta}(\Eb/\N)$, depending in general on $\Eb/\N$. At higher load, mutual information of TH is larger than spectral efficiency of DS. This analysis is intrinsically conservative, since spectral efficiency of TH will be, in general, larger than or equal to the mutual information obtained assuming Gaussian inputs. Therefore, one should expect that the gap in spectral efficiency between DS and TH with linear receivers is smaller and larger than that showed on figure when $\beta<\bar{\beta}$ and $\beta>\bar{\beta}$, respectively. 

\section{Conclusions}\label{sec:conc}

Verd\'u and Shamai showed in \cite{VerSha:1999} that optimum decoding provides a substantial gain over linear decoding in DS-CDMA, with random spreading. In particular, a bank of single-user matched filters followed by independent decoders is bounded in spectral efficiency at high-SNR, and linear multiuser detectors are needed in order to recover a nonzero spectral efficiency high-SNR slope. This behavior is partly due to the ``even'' use of degrees of freedom---coinciding in our setting with chips---that is intrisic of DS-CDMA \cite{MedGal:2002}.

The object of this paper was to analyze TH-CDMA with random hopping, and compare its behavior against DS-CDMA; we interpreted time-hopping in the general framework developed in \cite{VerSha:1999,ShaVer:2001}. The present analysis allowed comparison of TH vs. DS with same energy per symbol and same bandwidth constraints, and, therefore, showed the effect of the energy ``concentration,'' that is typical of TH. The degree of ``unevenness'' in TH-CDMA is directly related to the number of pulses $N_{\msf{s}}$ representing each symbol. At one extreme, one has maximum ``unevenness,'' where all energy is concentrated in one pulse ($N_{\msf{s}}=1$), while the other extreme corresponds to maximum ``evenness,'' $N_{\msf{s}}=N$, where TH coincides with DS. Particular emphasis has been put on the archetypal case of ``unevennes,'' that is $N_{\msf{s}}=1$, and partial results showing the general behavior when $N_\msf{s}>1$ have been derived.

A first result of our analysis was to derive a closed form expression for spectral efficiency of TH-CDMA with optimum decoding when $N_{\msf{s}}=1$, showing that, in this case, DS-CDMA outperforms TH-CDMA, in particular in the high-SNR regime. Same wideband behavior, but lower high-SNR slope, was observed for TH-CDMA vs. DS-CDMA, that is $\min\{1,\beta\}=\CMcal{S}_{\infty,\textup{DS}} > \CMcal{S}_{\infty,\textup{TH}}=1-e^{-\beta}$. A closed form expression for generic $N_{\msf{s}}$ remains an open problem; results based on simulations suggested, however, that the spectral efficiency loss at high-SNR may be considerably reduced while maintaining the number of pulses finite, and we provided evidences that the gap is reduced to a very small value with as low as two pulses per symbol ($N_{\msf{s}}=2$). This result indicates that the spectral efficiency gap may be substantially reduced while only using a fraction $N_{\msf{s}}/N$ of degrees of freedom \textit{per user}, that asymptotically vanishes as $N$ grows. 

A different behavior of TH-CDMA with respect to DS-CDMA was observed with linear receivers. Contrarily to DS, spectral efficiency of SUMF for TH with $N_{\msf{s}}=1$ was unbounded. As suggested, this asymptotic behavior may be traced back to the non-Gaussian distribution of the interference-plus-noise variable observed by each independent single-user decoder, that, in turn, depends on cross-correlation properties of spreading sequences. The same high-SNR slope $\CMcal{S}_{\infty,\textup{TH}}^{\msf{sumf}}=\beta e^{-\beta}$ was achieved by TH irrespectively of the knowledge that each single-user decoder had about spreading sequences of all other users. It was interesting to note that the maximum slope for TH, providing a hint on greatest energy efficiency, was reached when the number of users $K$ was equal to the number of chips $N$, \ie, $\beta=1$, leading to $\CMcal{S}_{\infty,\textup{TH}}^{\msf{sumf}}=1/e\approx 0.367879$. On the contrary, for $N_{\msf{s}}=\alpha N$, $\alpha\in(0,1)$, same spectral efficiency as DS-CDMA ($\alpha=1$) was obtained irrespectively of $\alpha$ for $N\to\infty$. 

The bounded nature of spectral efficiency with a SUMF bank in DS-CDMA is overcome, as well known, by using more complex linear receivers, that also account for interference, such as MMSE and decorrelator. Conversely, we showed that, in TH-CDMA, mutual information assuming Gaussian inputs has the same expression, irrespective of the linear receiver used, due to the peculiar structure of TH spreading sequences. TH sequences are indeed ``more'' likely to be linearly dependent than DS ones, in agreement with the intuition based on the cardinality of binary DS vs. TH codes, that is $2^{N}$ vs. $2N$. This lack of independence led to the impossibility of removing interference, which is instead almost surely feasible for DS, \eg with either decorrelator or MMSE receivers, as long as the load $\beta<1$. Therefore, in a low load, high-SNR scenario, DS outperforms TH. The opposite is true when $\beta>1$. In fact, while spectral efficiency in DS with MMSE rapidly drops, in particular with large $\Eb/\N$, as soon as $\beta$ becomes larger than one, mutual information of TH decays softly when one keeps overloading the system, and tends to the same MMSE DS limit. The absence of a spectral efficiency ``transition'' in the  neighborhood of the unit load, that is typical of DS, allows TH to outperform DS with any load larger than $\beta=1$ for sufficiently high $\Eb/\N$.

Beyond the natural extension of the present work to channels with fading, where the effect of an ``uneven'' use of degrees of freedom typical of TH should be investigated, we do stress that, from the single-user perspective, TH is a particular instance of impulsive signal. As such, the present theoretical setting, if appropriately adapted to asynchronous links, may serve as a basis for refining the understanding of the limits of impulsive communications.

\section*{Acknowledgment}
This work was partly supported by European Union and European Science Foundation under projects COST Action IC0902 ({Cognitive Radio and Networking for Cooperative Coexistence of Heterogeneous Wireless Networks}), and FP7 Network of Excellence ACROPOLIS ({Advanced coexistence technologies for radio optimization in licensed and unlicensed spectrum}), and by Sapienza University of Rome with the Research Grant, Anno 2013 - prot. C26A13ZYM2, and the ``Avvio alla Ricerca'' Project ICNET. The authors are grateful to two anonymous reviewers and the Editor for providing insightful suggestions and comments that allowed to significantly improve the paper and refined the understanding of the analyzed system models.

\appendices
\section{Proof of Theorem~\ref{theo:Nsonemomentsna}}
\label{app:dimfinite}

This Appendix is split in two parts. In the first part, we will find average moments $\E{{m}_L}$ for finite dimensional systems, where both $K$ and $N$ are finite. In the second part, we will prove that $\Var{m_L}\to0$, hence showing convergence in probability of $m_L$ to the $L$th moment of a Poisson distribution in the LSL.
%



\medskip
\textit{Part 1: Average Moments of TH-CDMA matrices with $N_\msf{s}=1$.}
\smallskip

Denote $\pi_k\in \bset{N}$ the nonzero element of the $k$th column $\bs{s}_k$ of $\bs{S}$. Then:
\[ \bs{S}\bs{S}^\t = \sum_{k=1}^K \bs{s}_k\bs{s}_k^\t = \sum_{k=1}^K \bs{e}_{\pi_k}\bs{e}_{\pi_k}^\t, \]
where $[\slim\bs{e}_i]_j=\delta_{ij}$, being $\delta_{ij}$ the Kronecker symbol. Hence, $\bs{S}\bs{S}^\t$ is diagonal, and the $n$th element on the diagonal, denoted $\nu_{nN}\eqdef [\bs{S}\bs{S}^\t]_{nn}$, is equal to:
\[ \nu_{nN} = |\{k\in\bset{K}\colon \pi_k=n\}| \in \bset{K}. \]

The $L$th moment of the ESD is:

%
\begin{equation} 
m_L = {\frac{1}{N} \Tr(\bs{S}\bs{S}^\t)^L} 
	 = \frac{1}{N}\sum_{i\leq N} [\bs{S}\bs{S}^\t]_{ii}^L 
	 = \frac{1}{N}\sum_{i\leq N} \nu_{iN}^L.
\end{equation}
Now note that $(\nu_{iN})_{i=1}^N$ is distributed according to a multinomial distribution with $\beta N$ trials and $N$ equally probable categories, that is, $(\nu_{1N},\dotsc,\nu_{NN})\thicksim\textup{Multinomial}(\beta N,N^{-1}\bs{1}_N)$. The marginal distribution of each $\nu_{iN}$ is $\textup{Binomial}(\beta N,N^{-1})$, thus:
\begin{align}
\E{m_L} 
	& = \frac{1}{N}\sum_{i\leq N} \Einline{\nu_{iN}^L} \nonumber \\
	& = \sum_{\ell=1}^{L} \left\{ \nslim{L \atop \ell} \right\} \frac{K!}{(K-\ell)!}\frac{1}{N^\ell}.  \label{eq:expand6}
\end{align}
In the LSL, one has:
\begin{equation}\label{eq:nonasymponeasympt} \E{m_L} \to \sum_{\ell=1}^L \left\{ \nslim{L \atop \ell} \right\} \beta^\ell, \end{equation}
that is exactly the Bell polynomial of order $L$, that provides the $L$th moment of a Poisson distribution with mean $\beta$.

\rem{Interestingly, the $L$th moment of the Mar\u{c}enko-Pastur law (c.f. eq.~\eqref{eq:mpdensity}) can be expressed as follows (see \eg \cite{BaiSil:1998, TulVer:2004}):
\begin{equation}\label{eq:mpmomexpl} m_{L}^{\msf{MP}} = \sum_{\ell=1}^{L} \CMcal{N}_{\ell} \beta^{\ell}, \quad \CMcal{N}_{\ell}=\frac{1}{L} \left({L \atop \ell}\right) \left({L \atop \ell-1}\right), \end{equation}
where $\CMcal{N}_{\ell}$ is the number of \textit{non-crossing partitions} of the set $\bset{L}$ into $\ell$ blocks, also known as \textit{Narayana number}. As a remark, the sum of Narayana numbers over $\bset{L}$ is the $L$th Catalan number, that has many combinatorial interpretations (see \eg \cite{Sta:2011,Sta:2001}). 

Also note that eq.~\eqref{eq:nonasymponeasympt} is formally similar to eq.~\eqref{eq:mpmomexpl}, with %
Stirling number of the second kind in place of Narayana numbers: While the latter enumerate non-crossing partitions only, the former enumerate all partitions, both crossing and non-crossing ones. 
%
}

\medskip
\textit{Part 2: $\Var{m_L}\to 0$.}
\smallskip
Exploiting the diagonal structure of $\bs{S}\bs{S}^\t$ yields the following expression for the second noncentral moment:
\begin{align} 
\Einline{m_L^2} 
	& = \frac{1}{N^2} \sum_{i\leq N}\sum_{j\leq N} \Einline{\nu_{iN}^L \nu_{jN}^L} \\
	& = \frac{1}{N} \Einline{\nu_{1N}^{2L}}+\Big(1-\frac{1}{N}\Big) \Einline{\nu_{1N}^L \nu_{2N}^L}.
\end{align}
The term $\Einline{\nu_{1N}^{2L}}=O(1)$, therefore $N^{-1}\Einline{\nu_{1N}^{2L}}$ contributes $O(N^{-1})$ to the sum. The term $\Einline{\nu_{1N}^L \nu_{2N}^L}$ can be handled as follows. Since the probability generating function (PGF) of the multinomial distribution describing $\bs{\nu}_N=(\nu_{1N},\dotsc,\nu_{NN})$ is:
\begin{equation}
G_{\bs{\nu}_N}(\bs{z}) = \Big( N^{-1}\sum_{m\leq N}z_m \Big)^{\!K},
\end{equation}
it follows that the PGF of the pair $(\nu_{1N},\nu_{2N})$ is:
\begin{equation}\begin{aligned}
G_{\nu_{1N}\nu_{2N}}(z_{1},z_{2}) 
	& = \Big( N^{-1}\Big(z_{1}+z_2+(N-2)\Big) \Big)^{\!K}\\
	& = e^{\beta(z_1-1)}e^{\beta(z_2-1)}+O(N^{-1}),
\end{aligned}\end{equation}
showing that $\nu_{iN}$ is asymptotically independent on $\nu_{jN}$, $i\neq j$, and distributed as a Poisson distribution with mean $\beta$. Therefore:
\begin{equation}
\Var{m_L} = \Einline{m_L^2}-\Einline{m_L}^2 = O(N^{-1}).
\end{equation}

\section{Verifying the Carleman condition}\label{app:carleman}
\begin{lemma} The sequence of moments $(\bar{m}_L)_{L\geq 0}$ %
verifies the Carleman condition, \ie, $\sum_{k\geq 1} \bar{m}_{2k}^{-1/2k}=\infty$.
\end{lemma}
\IEEEproof{We upper bound $\bar{m}_{2k}$ as follows:
\begin{align*}
\bar{m}_{2k} 
	& 	\overset{\textup{(a)}}{<} \sum_{\ell=1}^{2k} \left\{ {2k \atop \ell} \right\} \beta^\ell \\
	&	\overset{\textup{(b)}}{\leq} \sum_{\ell=1}^{2k-1}  \frac{1}{2} \left( {2k \atop \ell} \right) \ell^{2k-\ell} \beta^\ell+\beta^{2k} \\
	&	\overset{\textup{(c)}}{\leq} \frac{1}{2} (2k-1)^{2k-1} \sum_{\ell=1}^{2k-1}  \left( {2k \atop \ell} \right) \beta^\ell+\beta^{2k} \\
	&	\overset{\textup{(d)}}{<} \frac{1}{2} (2k-1)^{2k-1} (1+\beta)^{2k}+\beta^{2k} \\
	&	\overset{\textup{}}{<} (1+\beta)^{2k} (1+(2k)^{2k})
\end{align*}
where: (a) follows from the elementary inequality $(2k)!/(2k-\ell)!=(2k)(2k-1)\cdots(2k-\ell+1)<(2k)^\ell$; (b) from the inequality ${n \brace \ell}\leq U(n,\ell)\eqdef \frac{1}{2}\binom{n}{\ell}\ell^{n-\ell}$; (c) from upper bounding the term $\ell^{2k-\ell}$ with $(2k-1)^{2k-1}$; (d) from extending the summation over $\ell=0,\dotsc,2k$. %
From elementary relations between $p$-norms, one has $(1+(2k)^{2k})^{1/2k}=\|(1,2k)\|_{2k}\leq \|(1,2k)\|_1=1+2k$, thus $\bar{m}_{2k}^{1/2k} < (1+\beta)(1+2k)$, and therefore:
\[ \sum_{k\geq 1} \bar{m}_{2k}^{-1/2k} > \frac{1}{1+\beta} \sum_{k\geq 1} \frac{1}{1+2k}=\infty, \]
which verifies the Carleman condition.\hfill$\square$\medskip}

\section{Proof of $C^\msf{opt}_N(\gamma)\toprinline C^\msf{opt}(\gamma)$}\label{app:CntoC}
In this Appendix we show that $\Probinline{|C^\msf{opt}_N(\gamma)-C^\msf{opt}(\gamma)| > \epsilon} \to 0$, for all $\epsilon>0$. It is sufficient to prove that
\begin{equation}
\Var{A_N}\to 0, \quad A_N=\int h(x) d{\msf{F}^{\bs{S}\bs{S}^\t}_{\!N}\!}(x), \label{eq:vartoshow}
\end{equation}
when $h(x)$ is a concave, monotonically increasing function with sufficiently slow growth. (The case of interest is that of $h(x)$ with logarithmic growth.)
{

We use same notations as in Appendix~\ref{app:dimfinite}. Define the random measure:
\begin{equation}\label{eq:munu} \mu_N(x) = \frac{1}{N} \sum_{i=1}^N \indicator\{\nu_{iN}= x\}.
\end{equation}
The ESD ${\msf{F}^{\bs{S}\bs{S}^\t}_{\!N}\!}(x)$ is thus:
\begin{equation}\label{eq:Fnu} {\msf{F}^{\bs{S}\bs{S}^\t}_{\!N}\!}(x) = \mu_N([0,x]) = \sum_{k\leq x} \mu_{N}(k),
\end{equation}
hence $A_N$ in \eqref{eq:vartoshow} is
\begin{equation}
A_N=\int h(x) \,d{\msf{F}^{\bs{S}\bs{S}^\t}_{\!N}\!}(x) = \sum_{x\leq N} h(x) \mu_{N}(x).\label{eq:vartoshow2}
\end{equation}
The first moment of $A_N$ is:
 \begin{align}
\Einline{A_N} 
	& = \sum_{x\leq N} h(x) \Einline{\mu_{N}(x)} \\
	& = \sum_{x\leq N} h(x) \frac{1}{N}\sum_{i\leq N}\Einline{ \indicator\{\nu_{iN}= x\} } \\
	& = \sum_{x\leq N} h(x)\, \Probinline{\nu_{1N}= x}.
\end{align}
In order to compute the second moment of $A_N$ it is useful to preliminarily note that:
\begin{equation}\begin{aligned}
\sum_{i\leq N}\sum_{j\leq N}\Einline{ \indicator\{\nu_{iN}= x\}\indicator\{\nu_{jN}= x'\} } & = N\Probinline{ \nu_{1N}=x }\,\delta_{xx'} \\[-0.5ex]&+ N(N-1)\Probinline{ \{\nu_{1N}=x\}\cap\{\nu_{2N}=x'\} }.
\end{aligned}\end{equation}
For brevity, we denote $p_{1N}(x)=\Probinline{ \nu_{1N}=x }$ and $p_{2N}(x,x')=\Probinline{ \{\nu_{1N}=x\}\cap\{\nu_{2N}=x'\} }$. Hence, the second moment of $A_N$ is:
 \begin{align}
\Einline{A_N^2} 
	& = \sum_{x\leq N}\sum_{x'\leq N} h(x)h(x') \Einline{\mu_{N}(x)\mu_{N}(x')} \\
	& = \sum_{x\leq N}\sum_{x'\leq N} h(x)h(x') \frac{1}{N^2} \sum_{i\leq N}\sum_{j\leq N}\Einline{ \indicator\{\nu_{iN}= x\}\indicator\{\nu_{jN}= x'\} } \\
	& = \sum_{x\leq N}\sum_{x'\leq N} h(x)h(x') \bigg\{ \frac{1}{N}p_{1N}(x)\delta_{xx'} + \Big(1-\frac{1}{N}\Big)p_{2N}(x,x') \bigg\} .
\end{align}
Therefore, the LHS of \eqref{eq:vartoshow} is
\begin{align}
\Var{A_N} 
	& = \sum_{x\leq N}\sum_{x'\leq N} h(x)h(x') \bigg\{ \frac{1}{N}p_{1N}(x)\delta_{xx'} + \Big(1-\frac{1}{N}\Big)p_{2N}(x,x') - p_{1N}(x)p_{1N}(x') \bigg\} \\
	& \leq h(N)^2 \sum_{x\leq N}\sum_{x'\leq N} \frac{1}{N}p_{1N}(x)\delta_{xx'} + \Big(1-\frac{1}{N}\Big)p_{2N}(x,x') - p_{1N}(x)p_{1N}(x') \\
	& = h(N)^2 \sum_{x\leq N}\sum_{x'\leq N} \frac{1}{N}p_{1N}(x)\delta_{xx'} + \Big(1-\frac{1}{N}\Big)\bigg[ p_{1N}(x)p_{1N}(x')+O\Big(\frac{1}{N}\Big)\bigg] - p_{1N}(x)p_{1N}(x') \\
	& = h(N)^2 O\Big(\frac{1}{N}\Big).
\end{align}

\section{Asymptotics in the wideband regime for $N_\msf{s}>1$}
\label{app:dimNsg1wide}

\textit{1) Minimum energy-per-bit.}
\smallskip

We will show that:
Denote eigenvalues empirical average with $\bar{\lambda}_N$,
\[ \bar{\lambda}_N \eqdef \frac{1}{N}\sum_{i=1}^N \lambda_i. \]
Since
\begin{equation*}
\bar{\lambda}_N = \frac{1}{N}\sum_{i=1}^N \lambda_i = \frac{1}{N} \Tr(\bs{S}\bs{S}^\t) = \frac{1}{N} \Tr(\bs{S}^\t\bs{S}) 
= \frac{1}{N}\sum_{i=1}^K \bs{s}_i^\t \bs{s}_i =\frac{1}{N}\cdot K=\beta
\end{equation*}
surely, it results $\bar{\lambda}_N=\E{\lambda}=\beta$, hence 
%
the minimum energy-per-bit is:
\[ \eta_{\msf{min}} = \frac{\beta}{ \E{\lambda} }\ln2 = \ln 2. \]

\textit{2) Wideband slope $\CMcal{S}_0$.}
\smallskip

Denote the empirical second moment of eigenvalues with $S_N$,
\[ S_N\eqdef \frac{1}{N}\sum_{i=1}^N \lambda_i^2, \]
and let $R_{ij}=(\bs{S}^\t\bs{S})_{ij}$. Since:
\begin{equation}
 \frac{1}{N}\sum_{i=1}^N \lambda_i^2 = \frac{1}{N}\Tr(\bs{S}\bs{S}^\t)^2 = \frac{1}{N}\Tr(\bs{S}^\t\bs{S})^2 
 = \frac{1}{N} \sum_{i=1}^K \sum_{j=1}^K R_{ij}^2 = \beta+\frac{1}{N}\sum_{i=1}^K \sum_{\substack{j=1\\j\neq i}}^K R_{ij}^2, \label{eq:expanNs}
 \end{equation}
 one has:
 \[ \E{ S_N }=\beta+\frac{1}{N}K(K-1)\E{R_{12}^2}. \]
Denote
\[ \rho \eqdef R_{12}=\bs{s}_1^\t \bs{s}_2 = \sum_{m=1}^{N_\msf{s}} \bs{s}_{1m}^\t \bs{s}_{2m} = \sum_{m=1}^{N_\msf{s}} \rho_m, \]
 where $\bs{s}_{k}=[\bs{s}_{k1}^\t,\dotsc,\bs{s}_{kN_\msf{s}}^\t]^\t$ is an $(N_\msf{s},N_\msf{h})$-sequence (see Definition~\ref{def:THseq}) and $\rho_m\eqdef \bs{s}_{1m}^\t \bs{s}_{2m}$. The moment generating function (MGF) of $\rho$ is:
\begin{equation*} M_{\rho}(t)\eqdef \E{e^{t \rho}}=\E{e^{t(\rho_1+\cdots+\rho_{N_\msf{s}})}} = \big(\E{e^{t \rho_1 }}\big)^{N_\msf{s}} 
= \left({\int e^{t\xi} \mu(d\xi) }\right)^{\!\! N_{\msf{s}}},
\end{equation*}
where:
\[ \mu=\frac{1}{2N_\msf{h}}\delta_{-\frac{1}{N_\msf{s}}} + \Big(1-\frac{1}{N_\msf{h}}\Big)\delta_0 + \frac{1}{2N_\msf{h}}\delta_{\frac{1}{N_\msf{s}}}, \]
that is, explicitly,
\begin{equation}\label{eq:mgfx} M_{\rho}(t) = \left[ \Big({1-\frac{1}{N_\msf{h}}}\Big)+\frac{1}{N_\msf{h}} \cosh{\Big( \frac{t}{N_\msf{s}} \Big)} \right]^{N_\msf{s}}. \end{equation}
Hence $\E{\rho^2}=M_\rho''(0)=1/N$, and
\[ \Einline{ S_N }=\beta+\beta\frac{K-1}{N}\to\beta(1+\beta). \]
It is shown below that $\Var{S_N}=O(1/N)$, that implies $S_N\toprinline\beta+\beta^2$. From eq.~\eqref{eq:expanNs}, one has:
\begin{align*} \Einline{S_N^2} 
	& = \beta^2+\frac{2\beta}{N} K\cdot \frac{K-1}{N}+\frac{1}{N^2} {\sum_{i=1}^K\sum_{q=1}^K \sum_{\substack{j=1\\j\neq i}}^K\sum_{\substack{r=1\\r\neq q}}^K \Einline{R_{ij}^2 R_{qr}^2}} \\[-3pt]
	& = \beta^2+\frac{2\beta}{N} K\cdot \frac{K-1}{N}+\frac{1}{N^2} K(K-1)\Big(\Einline{\rho^4}+(K(K-1)-1)\Einline{\rho^2}^2 \Big)\\[4pt]
	& = \beta^2+2\beta^3 + \beta^4 + O{\Big(\frac{1}{N}\Big)}
\end{align*}
hence $\Varinline{S_N^2}=O(1/N)$.
\section{Mutual Information of SUMF when single-user decoders have knowledge on cross-correlations.}
\label{app:Nsleq1}

The SUMF channel for user $1$, as given by eq.~\eqref{eq:sumf2}, is:
\begin{equation*} 
\msf{y}_{1} = \msf{b}_1 + \sum_{k=2}^K \rho_{1k} \msf{b}_k + n_1. 
\end{equation*}
Assuming Gaussian inputs, $\msf{b}_i\sim\CNorm{0}{\En}$, the conditional mutual information on $\{\rho_{12},\dotsc,\rho_{1K}\}$ expressed in bits per channel use per user is: 
\begin{equation}\label{eq:Nsleq1_1}
 I(\msf{y}_1;\msf{b}_1| \rho_{12},\dotsc,\rho_{1K}) = \E{ \log_{2}\bigg( 1+\frac{\gamma}{1+\varsigma \gamma} \bigg) }, 
\end{equation}
where expectation is over $\{\rho_{12},\dotsc,\rho_{1K}\}$, and $\varsigma \eqdef \sum_{k=2}^K \rho_{1k}^2$. We find below the PDF of $\varsigma$ in the LSL.

From eq.~\eqref{eq:mgfx}, the characteristic function (CF) of the generic RV $\rho:=\rho_{1k}$ is: 
%
%
\begin{align} \varphi_\rho(t) 
 	 \eqdef \Einline{e^{it\rho}}= \left[1-\frac{N_\msf{s}}{N}\bigg(1-\cos\Big(\frac{t}{N_\msf{s}}\Big) \bigg)\right]^{N_\msf{s}} 
 	& =\ \sum_{m=0}^{N_\msf{s}} \left( {N_\msf{s} \atop m} \right) (-1)^m \frac{N_\msf{s}^m}{N^m} \bigg[1-\cos\Big(\frac{t}{N_\msf{s}}\Big) \bigg]^m \nonumber \\
& = 1 - \frac{N_\msf{s}^2}{N} \bigg[1-\cos\Big(\frac{t}{N_\msf{s}}\Big) \bigg] + O{\Big({\frac{1}{N^2}}\negthinspace\Big)}, \label{eq:phiX}
\end{align}
and the CF of $\varsigma$ is: 
\begin{equation}\label{eq:phisigma} \varphi_\varsigma(t)=\Einline{e^{it\varsigma}}=\Einline{e^{it\rho^2}}^{K-1}. \end{equation}
The last expectation can be computed as:
\[
\E{e^{it\rho^2}} 
	 = \frac{1}{2\pi} \int_\R d\omega\; \varphi_\rho(\omega) \int_\R dx\; e^{itx^2}  e^{-i\omega x} 
	 = \frac{1}{2\pi} \int_\R d\omega\; \varphi_\rho(\omega) \sqrt{i\frac{\pi}{t}} e^{-i\frac{\omega^2}{4t}} 
	 = 1+\frac{N_\msf{s}^2}{N}\Big[ e^{i{t}/{N_\msf{s}^2}}-1 \Big]+O{\Big({\frac{1}{N^2}}\!\Big)},
\]
hence, in the LSL, eq.~\eqref{eq:phisigma} becomes:
\[ \varphi_\varsigma(t) \to e^{\beta N_\msf{s}^2(e^{it/N_\msf{s}^2}-1)}, \]
which is the CF of a Poisson RV with measure:
\[ \mu_\varsigma = \sum_{k\geq 0} \frac{(\beta N_\msf{s}^2)^k}{k!}e^{-\beta N_\msf{s}^2}\delta_{k/N_\msf{s}^2}. \]
Therefore, from eq.~\eqref{eq:Nsleq1_1}, mutual information converges to:
\[  I(\msf{y}_1;\msf{b}_1| \rho_{12},\dotsc,\rho_{1K}) \to \sum_{k\geq0} \frac{(\beta N_\msf{s}^2)^{k}}{k!}e^{-\beta N_\msf{s}^2} \log_{2}{\negthinspace}\bigg(\negthinspace 1+\frac{\gamma}{1+\frac{k}{N_\msf{s}^2} \gamma} \bigg). \]


\section{Proof of Theorem~\ref{theo:SUMF} }
\label{app:SUMFproof}

Consider the output of the SUMF of user $1$, that is given by eq.~\eqref{eq:sumf2}, divided by $\sqrt{\N}$: 
\[Y_1	 = {b}_1 + \sum_{k=2}^K \rho_{1k} {b}_k + N_1
	 = {b}_{1} + J_1 + N_{1}
	 = {b}_{1} + Z_{1},
	\]
where $N_1\sim\CNorm{0}{1}$ and we assume ${b}_k\sim\CNorm{0}{\gamma}$. Since $I(Y_1;{b}_1)=h(Y_1)-h(Z_1)$, it is sufficient to find $P_{Y_1}$ and $P_{Z_1}$, both of which easily follow from $P_{J_1}$. %
%
%
%
From eq.~\eqref{eq:phiX}, one can write the CF of each term $\rho_{1k}{b}_k$ as:
\[ \varphi_{\rho{b}}(t)=\E{\varphi_\rho(bt)}=1-\frac{N_\msf{s}^2}{N}\left(1-e^{-\frac{\gamma}{2N_\msf{s}^2}t^2}\right)+O{\Big({\frac{1}{N^2}}\negthinspace\Big)}, \]
and, therefore, the CF of $J$ in the LSL is:
\begin{align*} \varphi_{J}(t) = \varphi_{\rho{b}}(t)^{K-1} 
	 \to \exp\Big[{\beta N_\msf{s}^2\Big( e^{-\frac{\gamma}{2N_\msf{s}^2}t^2}-1\Big)}\Big] 
	 = \sum_{k\geq 0} \frac{(\beta N_\msf{s}^2)^k}{k!} e^{-\beta N_\msf{s}^2} e^{-\frac{k\gamma}{2N_\msf{s}^2}t^2}, 
\end{align*}
which is the CF of:
\[ P_J = \sum_{k\geq 0} \frac{(\beta N_\msf{s}^2)^k}{k!} e^{-\beta N_\msf{s}^2} \CNorm{0}{k\gamma/N_\msf{s}^2}. \]
Therefore, $Z_1$ and $Y_1$ are distributed as:
\begin{equation} \label{eq:pdfZSUMF} P_Z = \sum_{k\geq 0} \frac{(\beta N_\msf{s}^2)^k}{k!} e^{-\beta N_\msf{s}^2} \CNorm{0}{1+k\gamma/N_\msf{s}^2}.  \end{equation}
and:
\begin{equation} \label{eq:pdfYSUMF} P_Y = \sum_{k\geq 0} \frac{(\beta N_\msf{s}^2)^k}{k!} e^{-\beta N_\msf{s}^2} \CNorm{0}{1+\gamma+k\gamma/N_\msf{s}^2}. \end{equation}

\section{Proof of Eq.~\eqref{eq:highSNRslopeSUMFTH} }
\label{app:SUMFhighSNRslopeproof}

The goal is to find the quantity $\CMcal{S}^{\,\msf{sumf}}_{\infty,\textup{TH}^\star} \eqdef \lim_{\gamma\uparrow\infty} \gamma {dI}/{d\gamma}$, where $I$ (nats/s/Hz) is given in Theorem~\ref{theo:SUMF}. Assuming that a limit does exist, we will upper and lower bound $h(P)$ with bounds having the same first derivative as $\gamma\to\infty$. Here $P$ is a generic linear combination of Gaussian distributions with weights $\msf{w}_k\geq 0$, as follows:
\[ P\eqdef \sum_{k\geq 0} \msf{w}_k\, \CNorm{\mu_k}{\sigma_{\!k}^2}. \]

\noindent\textit{Upper bound.} Applying the elementary inequality $(\msf{x}_1+\msf{x}_2)\ln(\msf{x}_1+\msf{x}_2) \geq \msf{x}_1\ln\msf{x}_1+\msf{x}_2\ln\msf{x}_2$ with $\msf{x}_1\geq 0$ and $\msf{x}_2\geq0$, properly generalized, %
to the differential entropy of $P$ yields:
\begin{align*}
h(P)  
	& = - \int_\C\! dz\, \Bigg( \sum_{k\geq 0} \msf{w}_{k}\, \CNorm{z;\mu_k}{\sigma_{\!k}^2}\! \Bigg) \, \ln \Bigg( \sum_{i\geq 0} \msf{w}_{i}\, \CNorm{z;\mu_i}{\sigma_{\!i}^2}\! \Bigg) \\
	& \leq  - \int_\C\! dz\, \sum_{k\geq 0} \msf{w}_{k}\, \CNorm{z;\mu_k}{\sigma_{\!k}^2} \ln{\negthinspace}\big[ \msf{w}_{k}\, \CNorm{z;\mu_k}{\sigma_{\!k}^2} \big] \\
	& = \sum_{k\geq0}-\msf{w}_k \ln\msf{w}_k + \sum_{k\geq0} \msf{w}_k\, \ln(\pi e{\sigma_{\!k}^2}) =: h_\msf{P}(P)+h_\msf{G}(P),
\end{align*}
%
%
%
%
where $h_\msf{P}(P)$ is constant in $\gamma$.

\noindent\textit{Lower bound.} From Gibb's inequality, 
\begin{align}
 h(P) 	 = \sum_{k\geq 0} \msf{w}_{k}\, \int_\C\! - {\slim}dz\,  \CNorm{z;\mu_k}{\sigma_{\!k}^2}  \, \ln\! \Bigg( \sum_{i\geq 0} \msf{w}_{i}\, \CNorm{z;\mu_i}{\sigma_{\!i}^2} \Bigg) 
		 \geq \sum_{k\geq 0}  \msf{w}_{k} \, \log(\pi e{\sigma_{\!k}^2}) =: h_\msf{G}(P).
\end{align}

\noindent\textit{Bounds.} From above bounds, it follows that $ h_\msf{G}(P) \leq h(P) \leq h_\msf{G}(P)+h_\msf{P}(P)$, and since $h_\msf{P}(P)$ is constant in $\gamma$, one has:
\begin{equation}\label{eq:gaussiantocompute}\lim_{\gamma\to\infty}\gamma\frac{dh(P)}{d\gamma} = \lim_{\gamma\to\infty}\gamma\frac{dh_\msf{G}(P)}{d\gamma}, \end{equation}
provided that the limit on the LHS does exist. Setting $\msf{w}_{k}\, \eqdef e^{-\beta}\beta^k/k!$ and $P$ equal to either $P_Y$ or $P_Z$ (c.f. eqs.~\eqref{eq:PYsumf}-\eqref{eq:PZsumf}) yields:
\begin{align}
\CMcal{S}_{\infty,\textup{TH}^\star}^{\msf{sumf}} 
	 = \beta\cdot \lim_{\gamma\to\infty} \gamma \frac{d}{d\gamma}\big\{ h_{\msf{G}}(P_Y)-h_{\msf{G}}(P_Z) \big\}; \label{eq:sthsumf}
\end{align}
by direct computations, it follows that:
\begin{align}
\frac{dh_\msf{G}(P_Z)}{d\gamma} & = \frac{1-e^{-\beta}}{\gamma}+O\Big({\frac{1}{\gamma^2}}\Big), \\
\frac{dh_\msf{G}(P_Y)}{d\gamma} & = \frac{1}{\gamma}+O\Big({\frac{1}{\gamma^2}}\Big),
\end{align}
hence $\CMcal{S}_{\infty,\textup{TH}^\star}^{\msf{sumf}}=\beta e^{-\beta}$. 



\section{Spectral Efficiency of SUMF for $N_{\msf{s}}=\alpha N$, $\alpha\in(0,1]$, As $N\to\infty$.}
\label{app:sumfcapacity}

In this appendix we will show that spectral efficiency of the single-user SUMF channel, given by eq.~\eqref{eq:sumf2}, 
is same as that of DS-CDMA for TH-CDMA with $N_{\msf{s}}=\alpha N$, $\alpha\in(0,1]$, therefore generalizing a previous result of Verd\'u and Shamai \cite{VerSha:1999} to which we reduce when $\alpha=1$. The result that follows extends the validity of the proof developed in \cite{VerSha:1999} to TH sequences with $N_\msf{s}=\alpha N$, $\alpha\in(0,1]$. The reader is referred to \cite{VerSha:1999} for a detailed exposition, while we limit the below derivation to our contribution, which reduces to verifying a Lindeberg-Feller condition for the interference term: 
\[ \lim_{K\to\infty} \sum_{k=2}^{K}\E{ \rho_{1k}^{2}\msf{b}_{k}^{2} \,\indicator{\{ \rho_{1k}^{2}\msf{b}_{k}^{2}>\xi \}} {\,}\big|{\,} \rho_{1k}} = 0, \quad \forall \xi>0, \]
%
%
that is equivalent to the following condition \cite{VerSha:1999}:
\begin{equation} \label{eq:suffcond}
\lim_{N\to\infty} \E{ N\rho_{12}^{2}\,\indicator{\{ N\rho_{12}^{2}>h \}}  } = 0,
\end{equation}
for some arbitrary $h>0$. Since the MGF of $\rho_{12}$ is given by eq.~\eqref{eq:mgfx}, one has, in the LSL with $N_\msf{s}=\alpha N$ and $N_h=N/N_\msf{s}=1/\alpha$, the following pointwise convergence of the MGF of $X\eqdef \sqrt{N}\rho$:
\[ M_{X}(t) = \Big[{1+\alpha\Big({\cosh\Big({\frac{t}{\sqrt{N}\alpha}}\Big)-1}\Big)}\Big]^{\alpha N} \!\!\longrightarrow\  e^{\frac{1}{2}t^2},\]
hence $X\todistrinline\Norm{0}{1}$. Therefore, as $h\to\infty$, it results $\Einline{ X^{2}\,\indicator{\{ X^{2}>h \}}  }\to 0$. %
%
%

\section{Closed form expression of eq.~\eqref{eq:genlin} for a General Class of Linear Receivers. }
\label{app:GENLINproof}



The discrete synchronous multiple-access channel considered is (c.f. eq.~\eqref{eq:discrsynch}):
\[\bs{y} = \bs{S}\bs{b}+\bs{n}. \]
The output of a generic linear receiver $\bs{W}^{\t}$ is as follows:
\begin{align} \bs{r} 	& \eqdef \bs{W}^{\t}\bs{y} 
				 = \bs{W}^\t \bs{S}\bs{b}+\bs{W}^\t\bs{n} \nonumber \\ 
				 & = \bs{G}\bs{b}+\bs{\nu}, \label{eq:defgenlin}
\end{align}
where $\bs{G}=\bs{W}^\t\bs{S}$ and $\bs{\nu}\sim\CNorm{\bs{0}}{\N\bs{W}^\t\bs{W}}$. We consider the following linear receiver structure parametrized by $\alpha$ and $\eta$:
\begin{equation} \label{eq:defW}
\bs{W}^{\t} = \bs{S}^\t(\eta \bs{S}\bs{S}^\t+\alpha \bs{I})^{-1};
\end{equation}
by setting $\eta=1$, decorrelator and MMSE receivers are obtained as special cases for $\alpha\to0$ and $\alpha=1/\gamma$, respectively; by setting $\eta=0$ and $\alpha\neq 0$, one obtains SUMF.

By focusing on user $1$, the output of channel of \eqref{eq:defgenlin} can be written as eq.~\eqref{eq:genlinuser1}, which is reported here for reference:
\begin{equation}\label{eq:genlinuser1} 
r_1 = G_{11} \msf{b}_1 + \sum_{k=2}^K G_{1k} \msf{b}_k + \nu_1. 
\end{equation}
%

As in the proof of Theorem~\ref{theo:Nsone}, we say that $s$ users are in chip $i$ when the $i$th diagonal element of $\bs{S}\bs{S}^\t$ is equal to $s$. Since $\bs{S}\bs{S}^\t$ is diagonal, one can write:
\[ \big[(\eta\bs{S}\bs{S}^\t+\alpha\bs{I})^{-1}\big]_{ii} = \frac{1}{\alpha+\eta u_i}, \]
where $u_i$ is the number of users in chip $i$; since $\bs{s}_i\in\{ \pm\bs{e}_n\}_{n=1}^N$, one has $\bs{s}_i=(-1)^{\msf{a}_i}\bs{e}_{\pi_i}$ for some $\msf{a}_i\in\{0,1\}$ and $\pi_i\in\bset{N}$, and therefore $u_i$ can be formally expressed as $u_i=|\{ k\in\bset{K}\colon \pi_k=i \}|$. The generic element $G_{ij}$ in eq.~\eqref{eq:genlinuser1} is explicitly given by:
\begin{align*}
G_{ij}  	
		& = \bs{s}_i^\t  (\eta \bs{S}\bs{S}^\t+\alpha \bs{I})^{-1} \bs{s}_j \\[3pt]
		& = (-1)^{\msf{a}_i} \bs{e}_{\pi_i}^\t (\eta \bs{S}\bs{S}^\t+\alpha \bs{I})^{-1} \bs{e}_{\pi_j} (-1)^{\msf{a}_j} \\
		& = (-1)^{\msf{a}_i+\msf{a}_j} \delta_{\pi_i,\pi_j} \cdot \frac{1}{\alpha+\eta u_{\pi_i}} 
		= \rho_{ij}\cdot \frac{1}{\alpha+\eta v_i},
\end{align*}
where we denoted by $v_i\eqdef u_{\pi_i}$, that is also equal to the number of spreading sequences equal to either $\bs{s}_i$ or $-\bs{s}_i$, \ie:
\[ v_{i} = \sum_{k=1}^{K} |\rho_{ik}| = \sum_{k=1}^{K} |\rho_{ki}|. \]
With similar computations, the generic element of the conditional covariance matrix of the noise vector in eq.~\eqref{eq:defgenlin} given $\{v_i\}$, and, therefore, given $\bs{S}$, is: 
\[ [\bs{\Sigma}_{\bs{\nu}|\bs{S}}]_{ij}= {\Einline{\bs{W}^\t\bs{n}\bs{n}^\herm\bs{W}\, \big| \, \bs{S}}}_{ij} = \N [\bs{W}^\t\bs{W}]_{ij} = \N \rho_{ij}\cdot \frac{1}{(\alpha+\eta v_i)^2}.\]

\normalsize
\ifCLASSOPTIONcaptionsoff
  \newpage
\fi



%
\bibliographystyle{IEEEtran}
\bibliography{biblioguido_tran_it_1_rev}

\end{document}